\pdfoutput=1
\documentclass[a4paper,11pt]{article}

\usepackage[
    margin=0.8in
    ]{geometry}

\usepackage{hyperref}
\hypersetup{
    colorlinks=true,
    linktocpage=true,
    linkcolor=blue,
    urlcolor=blue,
    citecolor=blue,
    filecolor=blue,
    bookmarksopen=true,
    }

\usepackage{amsmath,amssymb,amsthm,mathtools,mathrsfs}
\usepackage[italicdiff]{physics}
\mathtoolsset{showonlyrefs,showmanualtags}
\numberwithin{equation}{section}
\allowdisplaybreaks[4]

\usepackage[
    bibstyle=phys,
    sorting=none,
    biblabel=brackets,
    citestyle=numeric-comp,
    pageranges=false,
    eprint=true,
    url=true,
    ]{biblatex}
\DeclareFieldFormat{url}{[\href{#1}{iNSPIRE}]}
\DeclareSourcemap{
	\maps[datatype=bibtex]{
		\map[overwrite=true]{
			\step[fieldsource=shortjournal]
			\step[fieldset=journal, origfieldval]
		}
	}
}
\AtEveryBibitem{
    \clearfield{note}
    \clearname{editor}
    \clearlist{location}
    \iffieldequalstr{type}{NoiNSPIRE}{
        \DeclareFieldFormat{url}{[\href{#1}{Link}]}
        }{}
}
\bibliography{reference.bib}

\makeatletter
\renewcommand\thefootnote{\@fnsymbol\c@footnote}%
\renewcommand{\maketitle}{%
    \newpage
    \begin{flushright}
        UTHEP- 779
    \end{flushright}
    \null
    \vskip 2em%
    \begin{center}%
    \let \footnote \thanks
    {\LARGE\textbf{\@title}\par}%
    \vskip 1.5em%
    {\large
        \lineskip .5em%
        \begin{tabular}[t]{c}%
        \@author
    \end{tabular}\par}%
    \vskip 1em%
    {\large \@date}%
    \end{center}%
    \par
    \@thanks
    \vskip 1.5em}
\makeatother
\usepackage{authblk}

\title{Energy from Ellwood invariant for solutions involving $X^0$ variables}
\author{Yuji Ando\thanks{E-mail: \href{mailto:ando@het.ph.tsukuba.ac.jp}{ando@het.ph.tsukuba.ac.jp}}}
\author{Tomoya Suda\thanks{E-mail: \href{mailto:suda@het.ph.tsukuba.ac.jp}{suda@het.ph.tsukuba.ac.jp}}}
\affil{
Degree Programs in Pure and Applied Sciences,\\
Graduate School of Science and Technology, University of Tsukuba,\\
Tsukuba, Ibaraki 305-8571, Japan}
\date{}

\usepackage{tikz}

\begin{document}
\maketitle
\renewcommand\thefootnote{\arabic{footnote}}
\setcounter{footnote}{0}
\begin{abstract}
    For some classical solutions $\Psi_\mathrm{sol}$ in Witten's bosonic string field theory, it was proven that energy of the solution is proportional to the Ellwood invariant $\Tr(\mathcal{V}\Psi_\mathrm{sol})$ with $\mathcal{V}=c\bar{c}\partial X^0\bar{\partial}X^0$. We examine the relation for solutions involving $X^0$ variables. As a result, we obtain that the relation may not hold for such solutions. Namely, there is a possibility that the energy is not proportional to the Ellwood invariant.
\end{abstract}
\thispagestyle{empty}
\newpage
\setcounter{page}{1}
\tableofcontents
\section{Introduction}
String field theory has been actively studied as a candidate for a non-perturbative formulation of string theory. One of open bosonic string field theories is Witten's bosonic string field theory and the action is given by \cite{Witten:1985cc}
\begin{align}
    S[\Psi]=-\frac{1}{g^2}\Tr(\frac{1}{2}\Psi Q\Psi+\frac{1}{3}\Psi^3),
\end{align}
where $g$ is the coupling constant of the string field theory. In this theory, many classical solutions including tachyon vacuum solution were constructed, e.g. \cite{Schnabl:2005gv,Erler:2009uj,Fuchs:2008cc,Okawa:2012ica,Erler:2019vhl}.

To understand the physical interpretation of these solutions, it is important to compute physical observables. In the Witten's bosonic string field theory, two important observables exist. One is the energy of the classical solution. Because the action evaluated on a static solution is equal to minus the energy of the solution times the volume of the time coordinate, the energy of any static solution $\Psi_\mathrm{sol}$ is given by
\begin{align}
    E[\Psi_\mathrm{sol}]=-\frac{1}{\mathrm{Vol}(X^0)}S[\Psi_\mathrm{sol}].
\end{align}
Another is
\begin{align}
    \Tr(\mathcal{V}\Psi_\mathrm{sol}),
\end{align}
where $\mathcal{V}$ is a BRS invariant closed string state at the midpoint \cite{Gaiotto:2001ji,Ellwood:2008jh}. This is called Ellwood invariant. It is believed to be equal to the shift in the closed string tadpole amplitude between BCFTs described by the classical solution and the perturbative vacuum solution.

In \cite{Baba:2012cs}, they proved that the energy is proportional to the Ellwood invariant with
\begin{align}
    \mathcal{V}=\frac{2}{\pi i}c\bar{c}\partial X^0\bar{\partial}X^0.
\end{align}
However, it was shown for only some static classical solutions that do not involve $X^0$. Even if the solution involves $X^0$, the similar relation should hold as long as the solution is invariant under the shift of $X^0$ and it depends effectively only on derivatives of $X^0$. There is a possibility that solutions exist for which these conditions do not hold. In this paper, we examine the relation between the energy and the Ellwood invariant for static solutions that are constructed by $K,B,c$ and matter operators involving $X^0$. As a result, we obtain that there is a possibility that the energy is not proportional to the Ellwood invariant for such solutions.

This paper is organized as follows. In section \ref{review:KBc}, we review the discussion in \cite{Baba:2012cs} and confirm that the Ellwood invariant is proportional to the energy for regular solutions using only $K,B,c$. This includes not only Okawa type solution \cite{Okawa:2006vm} but also ghost brane solution \cite{Masuda:2012kt} and so on. In section \ref{original:KBcX}, we examine the relation for regular solutions which are constructed by $K,B,c$ and matter operators\footnote{In \cite{Baba:2012cs}, they prove for BMT solution \cite{Bonora:2010hi} which involves a relevant matter operator. We consider solutions that are constructed by not only the relevant operator but any matter operator involving $X^0$.}, and we obtain that there is a possibility that the energy is not proportional to the Ellwood invariant for such solutions. Additionally, we show the difference between the energy and the Ellwood invariant. In section \ref{summary}, we present the summary. Appendix \ref{Ape:Xcorr} gives formulas for correlation functions of the $X^\mu$ operators in sliver frame. In appendix \ref{appendixcond:3} and \ref{cond:1}, we examine relations that are needed to show that the energy is proportional to the Ellwood invariant.
\section{Review on Ellwood invariant and energy for \texorpdfstring{$KBc$}{KBc} solution\label{review:KBc}}
Many solutions are constructed by using string fields $K,B,c$. In this section, we consider string fields that are constructed only by $K,B,c$, and we call such solutions $KBc$ solutions.

The $KBc$ solutions can be written by
\begin{align}
    \Psi=\sum_i\sqrt{F_{1i}}c\frac{B}{H_i}c\sqrt{F_{2i}},
\end{align}
where $F_{1i},F_{2i}$ and $H_i$ are functions of $K$. As a concrete example, Okawa type solution \cite{Okawa:2006vm} is given by
\begin{align}
    \Psi=\sqrt{F_{11}}c\frac{B}{H_1}c\sqrt{F_{21}}\qc H_1=\frac{1-F_{11}}{K}\qand F_{11}=F_{21},
\end{align}
and ghost brane solution \cite{Masuda:2012kt} is given by
\begin{align}
    \Psi=\sqrt{F}_{11}c\frac{B}{H_1}c\sqrt{F_{12}}+\sqrt{F}_{21}c\frac{B}{H_2}c\sqrt{F_{22}}\qc H_i=\frac{1-F_{1i}}{K}\qand F_{i1}=F_{i2}\qand F_{12}F_{21}=1.
\end{align}
We represent $\sqrt{F_{ji}},1/H_i$ by a Laplace transform respectively:
\begin{align}
    \sqrt{F_{ji}}&\equiv\mathcal{L}\qty{f_{ji}}=\int_0^\infty\dd{L}e^{-LK}f_{ji}(L),\\
    \frac{1}{H_i}&\equiv\mathcal{L}\qty{h_i}=\int_0^\infty\dd{L}e^{-LK}h_i(L).
\end{align}
The $KBc$ solutions are also represented by the Laplace transform:
\begin{align}
    \begin{alignedat}{1}
        \Psi&=\int_0^\infty\dd{L}e^{-LK}\psi(L)\\
        &\psi(L)\coloneqq\sum_i\int\dd{L_{1i}dL_{2i}dL_{3i}}\delta(L-L_{1i}-L_{2i}-L_{3i})f_{1i}(L_{1i})h_i(L_{2i})f_{2i}(L_{3i})c(L_{2i}+L_{3i})Bc(L_{3i}),
    \end{alignedat}
\end{align}
where
\begin{align}
    c(z)\coloneqq e^{zK}ce^{-zK}.\label{notation:c}
\end{align}

Let us consider a test state $\Phi$
\begin{align}
    \Phi=e^{-\frac{1}{2}K}\phi e^{-\frac{1}{2}K},
\end{align}
where the string field $\phi$ is an infinitely thin strip with a boundary insertion of an operator $\phi(0)$. Similarly to \eqref{notation:c}, $\Phi$ can be also represented as
\begin{align}
    \Phi=e^{-K}\phi\qty(\frac{1}{2}).
\end{align}
Then the trace of $\Phi\Psi$ is given by the correlation function on the infinite cylinder
\begin{align}
    \Tr(\Phi\Psi)=\int_0^\infty\dd{L}\ev{\phi\qty(L+\frac{1}{2})\psi(L)}_{C_{L+1}},
\end{align}
where $C_{L+1}$ is the infinite cylinder with circumference $L+1$ and the map $f_2$ is defined \eqref{def:slivermap}. 

\begin{figure}
    \centering
    \begin{tikzpicture}
        \coordinate (r) at (2,0);
        \coordinate (l) at (-2,0);
        \draw[->] (1,0.1)--++(0,1);
        \draw[->] (1,2.1)--++(-1,0);
        \draw[->] (-1,2)--++(0,-1);
        \draw (1,0.1) node[above right] {$L+i\delta$}--++(0,2) node[right] {$L+i\Lambda$}--++(-2,0) node[left] {$i\Lambda$}--(-1,0.1) node[above left] {$i\delta$};
        \draw[thick] (r)--(l);
        \draw (2.5,2.5)--++(0,-1/2)--++(1/2,0);
        \node at (11/4,9/4) {$z$};
    \end{tikzpicture}
    \caption{The contour $P_{L,\Lambda,\delta}$}\label{figure:contourP}
\end{figure}
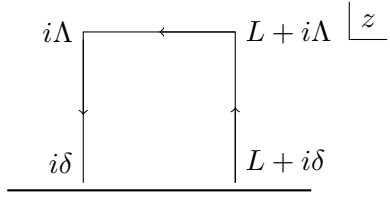

Let us consider $\mathcal{G}$ such that
\begin{align}
    \Tr(\Phi\mathcal{G}\Psi)=\lim_{(\Lambda,\delta)\to(\infty,0)}\int_0^\infty\dd{L}\ev{\phi\qty(L+\frac{1}{2})\mathcal{G}(L,\Lambda,\delta)\psi(L)}_{C_{L+1}},
\end{align}
where $\mathcal{G}(L,\Lambda,\delta)$ is defined by
\begin{align}
    \mathcal{G}(L,\Lambda,\delta)&\coloneqq\int_{P_{L,\Lambda,\delta}}\frac{\dd{z}}{2\pi i}g_z(z,\bar{z})-\int_{\bar{P}_{L,\Lambda,\delta}}\frac{\dd{\bar{z}}}{2\pi i}g_{\bar{z}}(z,\bar{z}),\\
    &g_z(z,\bar{z})\coloneqq2(X^0(z,\bar{z})-X^0(i\infty,-i\infty))\partial X^0(z),\\
    &g_{\bar{z}}(z,\bar{z})\coloneqq2(X^0(z,\bar{z})-X^0(i\infty,-i\infty))\bar{\partial}X^0(\bar{z}),
\end{align}
and the contour $P_{L,\Lambda,\delta}$ is in Figure \ref{figure:contourP} and the contour $\bar{P}_{L,\Lambda,\delta}$ is given in a similar way.
In \cite{Baba:2012cs}, they proven that $\mathcal{G}$ satisfies conditions
\begin{align}
    \frac{1}{2}\Tr(\Psi_\mathrm{sol}Q\Psi_\mathrm{sol})&=\Tr((\mathcal{G}\Psi_\mathrm{sol})Q\Psi_\mathrm{sol})-\frac{1}{2}\Tr(\Psi_\mathrm{sol}\comm{Q}{\mathcal{G}}\Psi_\mathrm{sol}),\label{cond:1a}\\
    \frac{1}{3}\Tr(\Psi_\mathrm{sol}^3)&=\Tr((\mathcal{G}\Psi_\mathrm{sol})\Psi_\mathrm{sol}^2),\label{cond:1b}\\
    \Tr(\Psi_\mathrm{sol}\comm{Q}{\mathcal{G}}\Psi_\mathrm{sol})&=\Tr(\Psi_\mathrm{sol}(\chi-\chi^\dag)\Psi_\mathrm{sol}),\label{cond:2}
\end{align}
for some static solutions $\Psi_\mathrm{sol}$, where $\chi$ is defied by
\begin{align}
    \chi\coloneqq\lim_{(\Lambda,\delta)\to(\infty,0)}\qty(\int_{i\delta}^{i\Lambda}\frac{\dd{z}}{2\pi i}4\partial X^0(z)\bar{c}\bar{\partial}X^0(\bar{z})-\int_{-i\delta}^{-i\Lambda}\frac{\dd{\bar{z}}}{2\pi i}4\bar{\partial}X^0(\bar{z})c\partial X^0(z)+\frac{c(0)}{2\pi\delta}).\label{def:chi}
\end{align}
Then the action evaluated on the solution can be written by
\begin{align}
    -S[\Psi_\mathrm{sol}]&=\Tr(\frac{1}{2}\Psi_\mathrm{sol}Q\Psi_\mathrm{sol}+\frac{1}{3}\Psi_\mathrm{sol}^3)\\
    &=\Tr(\mathcal{G}\Psi_\mathrm{sol}(Q\Psi_\mathrm{sol}+\Psi_\mathrm{sol}^2)-\frac{1}{2}\Psi_\mathrm{sol}\comm{Q}{\mathcal{G}}\Psi_\mathrm{sol})\\
    &=-\frac{1}{2}\Tr(\Psi_\mathrm{sol}(\chi-\chi^\dag)\Psi_\mathrm{sol})\\
    &=\Tr(\chi Q\Psi_\mathrm{sol})\\
    &=\Tr(\qty(\frac{2}{\pi i}c\bar{c}\partial X^0\bar{\partial}X^0)\Psi_\mathrm{sol}).
\end{align}
Here if the following holds
\begin{align}
    \Tr(Q\chi\Psi_\mathrm{sol})=\Tr(\qty(\frac{2}{\pi i}c\bar{c}\partial X^0\bar{\partial}X^0)\Psi_\mathrm{sol}),\label{cond:3}
\end{align}
the energy is proportional to the Ellwood invariant.

In this section, we consider
\begin{align}
    \Tr(\Psi\comm{Q}{\mathcal{G}}\Psi)=\mathcal{A}_1+\mathcal{A}_2,
\end{align}
where
\begin{align}
    \mathcal{A}_1&\coloneqq\lim_{(\Lambda,\delta)\to(\infty,0)}\int_0^\infty\dd{L'}\int_0^\infty\dd{L}\ev{\psi(L+L')\comm{Q}{\mathcal{G}(L,\Lambda,\delta)}\psi(L)}_{C_{L+L'}},\\
    \mathcal{A}_2&\coloneqq\lim_{(\Lambda,\delta)\to(\infty,0)}\int_0^\infty\dd{L'}\int_0^\infty\dd{L}\ev{\psi(L+L')\mathcal{G}(L,\Lambda,\delta)\qty(Q\psi(L)-\mathcal{L}^{-1}\qty{Q\Psi}(L))}_{C_{L+L'}},
\end{align}
and we confirm
\begin{align}
    \mathcal{A}_1+\mathcal{A}_2=\Tr(\Phi(\chi-\chi^\dag)\Psi).
\end{align}
Here in appendix \ref{cond:1}, we check that \eqref{cond:1a} and \eqref{cond:1b} holds.
\subsection{Evaluation of \texorpdfstring{$\mathcal{A}_1$}{A1}}
Because of
\begin{align}
    &\comm{Q}{\mathcal{G}(L,\Lambda,\delta)}=\mathcal{C}_\mathrm{I}(L,\Lambda,\delta)+\mathcal{C}_{\mathrm{I}\hspace{-1.2pt}\mathrm{I}\mathrm{A}}(L,\Lambda,\delta)+\mathcal{C}_{\mathrm{I}\hspace{-1.2pt}\mathrm{I}\mathrm{B}}(L,\Lambda,\delta)\\
    &\qquad\mathcal{C}_\mathrm{I}(L,\Lambda,\delta)\coloneqq\int_{P_{L,\Lambda,\delta}}\frac{\dd{z}}{2\pi i}4\partial X^0(z)\bar{c}\bar{\partial}X^0(\bar{z})-\int_{\bar{P}_{L,\Lambda,\delta}}\frac{\dd{\bar{z}}}{2\pi i}4\bar{\partial}X^0(\bar{z})c\partial X^0(z),\\
    &\qquad\mathcal{C}_{\mathrm{I}\hspace{-1.2pt}\mathrm{I}\mathrm{A}}(L,\Lambda,\delta)\coloneqq-2\qty(c\partial X^0(i\infty)+\bar{c}\bar{\partial}X^0(-i\infty))\qty(\int_{P_{L,\Lambda,\delta}}\frac{\dd{z}}{2\pi i}\partial X^0(z)-\int_{\bar{P}_{L,\Lambda,\delta}}\frac{\dd{\bar{z}}}{2\pi i}\bar{\partial}X^0(\bar{z})),\\
    &\qquad\mathcal{C}_{\mathrm{I}\hspace{-1.2pt}\mathrm{I}\mathrm{B}}(L,\Lambda,\delta)\coloneqq\int_{P_{L,\Lambda,\delta}}\frac{\dd{z}}{2\pi i}\frac{1}{2}\partial^{2}c-\int_{\bar{P}_{L,\Lambda,\delta}}\frac{\dd{\bar{z}}}{2\pi i}\frac{1}{2}\bar{\partial}^{2}\bar{c}+\int_{P_{L,\Lambda,\delta}}\dd{z}\partial\kappa(z,\bar{z})+\int_{\bar{P}_{L,\Lambda,\delta}}\dd{\bar{z}}\bar{\partial}\kappa(z,\bar{z}),
\end{align}
with
\begin{align}
    \kappa(z,\bar{z})\coloneqq\frac{1}{2\pi i}\qty(c(z)g_z(z,\bar{z})-\bar{c}(\bar{z})g_{\bar{z}}(z,\bar{z})),
\end{align}
$\mathcal{A}_1$ is given by\footnote{We follow the notation in \cite{Hata:2013hba}.}
\begin{align}
    \mathcal{A}_1&=\mathcal{T}_\mathrm{I}+\mathcal{T}_{\mathrm{I}\hspace{-1.2pt}\mathrm{I}\mathrm{A}}+\mathcal{T}_{\mathrm{I}\hspace{-1.2pt}\mathrm{I}\mathrm{B}},\\
    &\mathcal{T}_\mathrm{I}\coloneqq\lim_{(\Lambda,\delta)\to(\infty,0)}\int_0^\infty\dd{L'}\int_0^\infty\dd{L}\ev{\psi(L+L')\mathcal{C}_\mathrm{I}(L,\Lambda,\delta)\psi(L)}_{C_{L+L'}},\\
    &\mathcal{T}_{\mathrm{I}\hspace{-1.2pt}\mathrm{I}\mathrm{A}}\coloneqq\lim_{(\Lambda,\delta)\to(\infty,0)}\int_0^\infty\dd{L'}\int_0^\infty\dd{L}\ev{\psi(L+L')\mathcal{C}_{\mathrm{I}\hspace{-1.2pt}\mathrm{I}\mathrm{A}}(L,\Lambda,\delta)\psi(L)}_{C_{L+L'}},\\
    &\mathcal{T}_{\mathrm{I}\hspace{-1.2pt}\mathrm{I}\mathrm{B}}\coloneqq\lim_{(\Lambda,\delta)\to(\infty,0)}\int_0^\infty\dd{L'}\int_0^\infty\dd{L}\ev{\psi(L+L')\mathcal{C}_{\mathrm{I}\hspace{-1.2pt}\mathrm{I}\mathrm{B}}(L,\Lambda,\delta)\psi(L)}_{C_{L+L'}}.
\end{align}

First, we evaluate $\mathcal{T}_\mathrm{I}$. Because $\psi(L)$ does not involve $X^0$, $\mathcal{T}_\mathrm{I}$ is factorized as
\begin{align}
    &\int_0^\infty\dd{L'}\int_0^\infty\dd{L}\ev{\psi(L+L')\mathcal{C}_\mathrm{I}(L,\Lambda,\delta)\psi(L)}_{C_{L+1}}\\
    &=4\int_0^\infty\dd{L'}\int_0^\infty\dd{L}\int_{P_{L,\Lambda,\delta}}\frac{\dd{z}}{2\pi i}\ev{\partial X^0(z)\bar{\partial}X^0(\bar{z})}_{C_{L+1}}\ev{\psi(L+L')\bar{c}(\bar{z})\psi(L)}_{C_{L+L'}}\\
    &\qquad-4\int_0^\infty\dd{L'}\int_0^\infty\dd{L}\int_{\bar{P}_{L,\Lambda,\delta}}\frac{\dd{\bar{z}}}{2\pi i}\ev{\bar{\partial}X^0(\bar{z})\partial X^0(z)}_{C_{L+1}}\ev{\psi(L+L')c(z)\psi(L)}_{C_{L+L'}}.\label{C1}
\end{align}
Here since for $y\to\infty$ with $z=x+iy$,
\begin{align}
    \ev{\partial X^0(z)\bar{\partial}X^0(\bar{z})}_{C_{L+L'}}&=-2\qty(\frac{\pi}{L+L'})^2e^{-4\pi y/(L+L')}+\order{e^{-8\pi y/(L+L')}},\\
    c(z)&\propto e^{\pi y/(L+L')},
\end{align}
the horizontal part of the contours $P_{L,\Lambda,\delta},\bar{P}_{L,\Lambda,\delta}$ does not contribute $\mathcal{T}_\mathrm{I}$ in the limit $\Lambda\to\infty$. Thus we obtain
\begin{align}
    \mathcal{T}_\mathrm{I}&=\lim_{(\Lambda,\delta)\to(\infty,0)}\left[\ev{\psi(L+L')\int_{i\delta}^{i\Lambda}\frac{\dd{z}}{2\pi i}4\partial X^0(z)\bar{c}\bar{\partial}X^0(\bar{z})\psi(L)}_{C_{L+L'}}\right.\\
    &\qquad\qquad\qquad+\ev{\psi(L+L')\psi(L)\int_{i\delta}^{i\Lambda}\frac{\dd{z}}{2\pi i}4\partial X^0(z)\bar{c}\bar{\partial}X^0(\bar{z})}_{C_{L+L'}}\\
    &\qquad\qquad\qquad+\ev{\psi(L+L')\int_{-i\delta}^{-i\Lambda}\frac{\dd{\bar{z}}}{2\pi i}4\bar{\partial}X^0(\bar{z})\partial X^0(z)\psi(L)}_{C_{L+L'}}\\
    &\qquad\qquad\qquad\left.+\ev{\psi(L+L')\psi(L)\int_{-i\delta}^{-i\Lambda}\frac{\dd{\bar{z}}}{2\pi i}4\bar{\partial}X^0(\bar{z})\partial X^0(z)}_{C_{L+L'}}\right].
\end{align}

Next, we evaluate $\mathcal{T}_{\mathrm{I}\hspace{-1.2pt}\mathrm{I}\mathrm{A}}$.
\begin{align}
    &\ev{\psi(L+L')\mathcal{C}_{\mathrm{I}\hspace{-1.2pt}\mathrm{I}\mathrm{A}}(L,\Lambda,\delta)\psi(L)}_{C_{L+L'}}\\
    &=-2\ev{\psi(L+L')\qty(c\partial X^0(i\infty)+\bar{c}\bar{\partial}X^0(-i\infty))\qty(\int_{P_{L,\Lambda,\delta}}\frac{\dd{z}}{2\pi i}\partial X^0(z)-\int_{\bar{P}_{L,\Lambda,\delta}}\frac{\dd{\bar{z}}}{2\pi i}\bar{\partial}X^0(\bar{z}))\psi(L)}_{C_{L+L'}}\\
    &=-2\left\langle \psi(L+L')\qty(c\partial X^0(i\infty)+\bar{c}\bar{\partial}X^0(-i\infty))\right.\\
    &\qquad\times\left.\frac{1}{2\pi i}\qty((X^0(L+i\delta)-X^0(L-i\delta))-(X^0(i\delta)-X^0(-i\delta)))\psi(L)\right\rangle_{C_{L+L'}}
\end{align}
Here we assume the boundary condition for $X^\mu$.
\begin{align}
    \partial X^\mu(z)=\bar{\partial}X^\mu(\bar{z})\qfor z=\bar{z}\label{b.c.:X}
\end{align}
Because of the boundary condition, the right-hand side of the above equation vanishes in the limit $\delta\to0$. Thus we obtain
\begin{align}
    \mathcal{T}_{\mathrm{I}\hspace{-1.2pt}\mathrm{I}\mathrm{A}}=0.
\end{align}

Finally, we evaluate $\mathcal{T}_{\mathrm{I}\hspace{-1.2pt}\mathrm{I}\mathrm{B}}$. We need to consider the anticommutator between $B$ and $\kappa$
\begin{align}
    \acomm{B}{\kappa(z,\bar{z})}=\frac{1}{2\pi i}\qty(g_z(z,\bar{z})-g_{\bar{z}}(z,\bar{z})),
\end{align}
because the contours $P_{L,\Lambda,\delta}$ and $\bar{P}_{L,\Lambda,\delta}$ cross $B$. Using this, anticommutator between $\psi$ and $\kappa$ is given by
\begin{align}
    \acomm{\psi(L)}{\kappa(z,\bar{z})}&=\sum_i\int\dd{L_{1i}dL_{2i}dL_{3i}}\delta(L-L_{1i}-L_{2i}-L_{3i})f_{1i}(L_{1i})h_i(L_{2i}f_{2i}(L_{3i})\\
    &\qquad\times c(L_{2i}+L_{3i})\frac{1}{2\pi i}\qty(g_z(a+i\Lambda,a-i\Lambda)-g_{\bar{z}}(a+i\Lambda,a-i\Lambda))c(L_{3i}),
\end{align}
where $L_{2i}+L_{3i}>a>L_{3i}$. Using \eqref{correlation:g}, we obtain
\begin{align}
    \ev{\psi(L+L')\acomm{\psi(L)}{\kappa(z,\bar{z})}}_{C_{L+L'}}=\frac{1}{L+1}\coth\frac{2\pi\Lambda}{L+1}\ev{\psi(L+L')\alpha(L)}_{C_{L+L'}},\label{formula:kappafactorize}
\end{align}
where $\alpha$ is defined by
\begin{align}
    \alpha(L)\coloneqq\sum_i\alpha_i(L)\qc\alpha_i(L)\coloneqq\mathcal{L}^{-1}\qty{\sqrt{F_{1i}}c\frac{1}{H_i}c\sqrt{F_{2i}}}.
\end{align}
Hence we obtain
\begin{align}
    &\ev{\psi(L+L')\mathcal{C}_{\mathrm{I}\hspace{-1.2pt}\mathrm{I}\mathrm{B}}(L,\Lambda,\delta)\psi(L)}_{C_{L+L'}}\\
    &=-\ev{\psi(L+L')\psi(L)\qty(\frac{1}{4\pi i}(\partial c(i\delta)-\bar{\partial}c(-i\delta))+\kappa(i\delta,-i\delta))}_{C_{L+L'}}\\
    &\qquad+\ev{\psi(L+L')\qty(\frac{1}{4\pi i}(\partial c(L+i\delta)-\bar{\partial}c(L-i\delta))+\kappa(L+i\delta,L-i\delta))\psi(L))}_{C_{L+L'}}\\
    &\qquad+\frac{1}{L+1}\coth\frac{2\pi\Lambda}{L+1}\ev{\psi(L+L')\alpha(L)}_{C_{L+L'}}.
\end{align}
Since for the boundary condition of $c$
\begin{align}
    c(z)=\bar{c}(\bar{z})\qfor z=\bar{z},\label{b.c.:c}
\end{align}
in the limit $\delta\to0$, the below equations hold.
\begin{align}
    \lim_{\delta\to0}\partial c(i\delta)&=\lim_{\delta\to0}\bar{\partial}c(-i\delta)\\
    \lim_{\delta\to0}\partial c(L+i\delta)&=\lim_{\delta\to0}\bar{\partial}c(L-i\delta)
\end{align}
In addition, using \eqref{correlation:g}, we can show
\begin{align}
    \lim_{\delta\to0}\ev{\kappa(i\delta,-i\delta)}_{C_{L+L'}}&=\frac{1}{2\pi i}\lim_{\delta\to0}\qty(\ev{g_z(i\delta,-i\delta)}_{C_{L+L'}}\ev{c(i\delta)}_{C_{L+L'}}-\ev{g_{\bar{z}}(i\delta,-i\delta)}_{C_{L+L'}}\ev{\bar{c}(-i\delta)}_{C_{L+L'}})\\
    &=\lim_{\delta\to0}\frac{1}{2\pi\delta}\qty(\ev{c(i\delta)}_{C_{L+L'}}-\ev{\bar{c}(-i\delta)}_{C_{L+L'}}).
\end{align}

Thus $\mathcal{T}_{\mathrm{I}\hspace{-1.2pt}\mathrm{I}\mathrm{B}}$ is
\begin{align}
    &\lim_{(\Lambda,\delta)\to(\infty,0)}\ev{\psi(L+L')\mathcal{C}_{\mathrm{I}\hspace{-1.2pt}\mathrm{I}\mathrm{B}}(L,\Lambda,\delta)\psi(L)}_{C_{L+L'}}\\
    &=\ev{\psi(L+L')\lim_{\delta\to0}\frac{1}{2\pi\delta}\qty(\qty(c(L+i\delta)-\bar{c}(L-i\delta))\psi(L)-\psi(L)\qty(c(i\delta)-\bar{c}(-i\delta)))}_{C_{L+L'}}\\
    &\qquad+\frac{1}{L+1}\ev{\psi(L+L')\alpha(L)}_{C_{L+L'}}
\end{align}

From our computation, we obtain
\begin{align}
    \mathcal{A}_1=\int_0^\infty\dd{L'}\int_0^\infty\dd{L}\Tr(e^{-L'K}\psi(L')\qty(\chi e^{-LK}\psi(L)+e^{-LK}\psi(L)\chi+\frac{1}{1+L}e^{-LK}\alpha(L))).
\end{align}
\subsection{Evaluation of \texorpdfstring{$\mathcal{A}_2$}{A2}\label{subsec:A2KBc}}
We evaluate $\mathcal{A}_2$. Here we note
\begin{align}
    \mathcal{L}^{-1}\qty{Q\Psi}-Q\mathcal{L}^{-1}\qty{\Psi}(L)=e^{LK}\partial_L(e^{-LK}\alpha(L))-\delta(L)\alpha(0).
\end{align}
See appendix B of \cite{Baba:2012cs} for the derivation. Using it and $\mathcal{G}(0,\Lambda,\delta)=0$, we obtain the below.
\begin{align}
    \ev{\psi(L+L')\mathcal{G}(L,\Lambda,\delta)\qty(Q\psi(L)-\mathcal{L}^{-1}\qty{Q\Psi}(L))}_{C_{L+L'}}&=\ev{\psi(L+L')\mathcal{G}(L,\Lambda,\delta)e^{LK}\partial_L(e^{-LK}\alpha(L))}_{C_{L+L'}}\\
    &=\eval{\partial_t\ev{\psi(L+L')\mathcal{G}(L,\Lambda,\delta)e^{-tK}\alpha(L+t)}_{C_{L+L'+t}}}_{t=0}
\end{align}
Since this can be factorized as
\begin{align}
    &\eval{\partial_t\ev{\psi(L+L')\mathcal{G}(L,\Lambda,\delta)e^{-tK}\alpha(L+t)}_{C_{L+L'+t}}}_{t=0}\\
    &=\eval{\partial_t\qty(\ev{\mathcal{G}(L,\Lambda,\delta)}_{C_{L+L'+t}}\ev{\psi(L+L')e^{-tK}\alpha(L+t)}_{C_{L+L'+t}})}_{t=0}\\
    &=\ev{\mathcal{G}(L,\Lambda,\delta)}_{C_{L+L'}}\ev{\psi(L+L')e^{LK}\partial_L\qty(e^{-LK}\alpha(L))}_{C_{L+L'}}\\
    &\qquad+\eval{\partial_t\ev{\mathcal{G}(L,\Lambda,\delta)}_{C_{L+L'+t}}}_{t=0}\ev{\psi(L+L')\alpha(L))}_{C_{L+L'}},\label{factorize:A2}
\end{align}
$\mathcal{A}_2$ can be written by
\begin{align}
    &\mathcal{A}_2\\
    &=\int_0^\infty\dd{L'}\int_0^\infty\dd{L}\lim_{(\Lambda,\delta)\to(\infty,0)}\ev{\mathcal{G}(L,\Lambda,\delta)}_{C_{L+L'}}\Tr(e^{-L'K}\psi(L')\partial_L(e^{-LK}\alpha(L)))\\
    &\qquad+\int_0^\infty\dd{L'}\int_0^\infty\dd{L}\lim_{(\Lambda,\delta)\to(\infty,0)}\eval{\partial_t\ev{\mathcal{G}(L,\Lambda,\delta)}_{C_{L+L'+t}}}_{t=0}\Tr(e^{-L'K}\psi(L')e^{-LK}\alpha(L))\\
    &=\int_0^\infty\dd{L'}\int_0^\infty\dd{L}\lim_{(\Lambda,\delta)\to(\infty,0)}\qty(\eval{\partial_t\ev{\mathcal{G}(L,\Lambda,\delta)}_{C_{L+L'+t}}}_{t=0}-\partial_L\ev{\mathcal{G}(L,\Lambda,\delta)}_{C_{L+L'}})\Tr(e^{-L'K}\psi(L')e^{-LK}\alpha(L))\\
    &\qquad+\int_0^\infty\dd{L'}\lim_{(\Lambda,\delta)\to(\infty,0)}\eval{\ev{\mathcal{G}(L,\Lambda,\delta)}_{C_{L+L'}}\Tr(e^{-L'K}\psi(L')e^{-LK}\alpha(L))}_{L=0}^{L=\infty}.
\end{align}
With the help of \eqref{correlation:G}, we can derive the following.
\begin{align}
    \lim_{(\Lambda,\delta)\to(\infty,0)}\qty(\eval{\partial_t\ev{\mathcal{G}(L,\Lambda,\delta)}_{C_{L+L'+t}}}_{t=0}-\partial_L\ev{\mathcal{G}(L,\Lambda,\delta)}_{C_{L+L'}})=-\frac{1}{L+1}
\end{align}
Using it and the assumption $\alpha(\infty)=0$, we obtain
\begin{align}
    \mathcal{A}_2=-\int_0^\infty\dd{L'}\int_0^\infty\dd{L}\frac{1}{1+L}\Tr(e^{-L'K}\psi(L')e^{-LK}\alpha(L)).
\end{align}

Therefore we can confirm that \eqref{cond:2} holds for the $KBc$ solutions which satisfy the assumption $\alpha(\infty)=0$
\begin{align}
    \mathcal{A}_1+\mathcal{A}_2=\int_0^\infty\dd{L'}\int_0^\infty\dd{L}\Tr(e^{-L'K}\psi(L')\qty(\chi e^{-LK}\psi(L)+e^{-LK}\psi(L)\chi)).
\end{align}
Because we expect $\alpha(\infty)=0$ for regular solutions, the energy of the regular $KBc$ solutions are proportional to the Ellwood invariant.
\section{Ellwood invariant and energy for the solutions including \texorpdfstring{$X^0$}{X0}\label{original:KBcX}}
Various solutions are constructed by not only $K,B,c$ but also string fields involving matter operators \cite{Schnabl:2007az,Kiermaier:2007ba,Kiermaier:2010cf,Bonora:2010hi,Erler:2014eqa}. Especially we focus on
\begin{align}
    \Psi=\sum_i\sqrt{F_{1i}}c\sqrt{F_{2i}}G_{1i}\frac{B}{H_i}G_{2i}\sqrt{F_{3i}}c\sqrt{F_{4i}},\label{def:KBcmatterstring}
\end{align}
where $G_{1i},G_{2i}$ are functions of string fields which are an infinitely thin strip with a boundary insertion of matter operators. In this case also, $F_{ji}$ and $H_i$ are represented by Laplace transform respectively. Then the solutions can be written by
\begin{align}
    \Psi&=\int_0^\infty\dd{L}e^{-LK}\psi(L),\\
    &\psi(L)\coloneqq\sum_i\int\dd{L_{1i}dL_{2i}dL_{3i}dL_{4i}dL_{5i}}\delta(L-L_{1i}-L_{2i}-L_{3i}-L_{4i}-L_{5i})\\
    &\qquad\qquad\times f_{1i}(L_{1i})f_{2i}(L_{2i})h_i(L_{3i})f_{3i}(L_{4i})f_{4i}(L_{5i})\\
    &\qquad\qquad\times c(L_{2i}+L_{3i}+L_{4i}+L_{5i})G_{1i}(L_{3i}+L_{4i}+L_{5i})BG_{2i}(L_{4i}+L_{5i})c(L_{5i}).
\end{align}
As a concrete example, simple intertwining solution \cite{Erler:2014eqa,Erler:2019vhl} is given by
\begin{gather}
    \Psi_\mathrm{int}=\sqrt{F_{11}}c\sqrt{F_{21}}G_{11}\frac{B}{H_1}G_{21}\sqrt{F_{21}}c\sqrt{F_{11}}+\sqrt{F_{12}}c\sqrt{F_{22}}G_{12}\frac{B}{H_2}G_{22}\sqrt{F_{22}}c\sqrt{F_{12}},\label{simpleintertwining}\\
    G_{11}=G_{21}=F_{21}=1\qc F_{11}=F_{12}=H_1=\frac{1}{1+K}\qc F_{22}=(1+K)^2\qc H_2=-(1+K),\\
    G_{12}=\sigma\qc G_{22}=\bar{\sigma},\label{simpleintertwining:coefficient}
\end{gather}
where $\sigma,\bar{\sigma}$ are defined as an infinitesimally thin strip with the respectively operators insertion by
\begin{align}
    \sigma=\sigma_*e^{i\sqrt{h}X^0}\qc\bar{\sigma}=\bar{\sigma}_*e^{-i\sqrt{h}X^0},\label{def:bcc}
\end{align}
and $\sigma_*,\bar{\sigma}_*$ are boundary condition changing operators and both of them are primaries of weight $h$.\footnote{In \cite{Erler:2019fye}, a flag state solution was constructed. This does not involve $X^0$, and the simple intertwining solution is derived as a limiting case of it. However, in this paper, we do not consider it.}

In this section, we study
\begin{align}
    \mathcal{A}_1+\mathcal{A}_2=\Tr(\Psi(\chi-\chi^\dag)\Psi),
\end{align}
for the solution \eqref{def:KBcmatterstring}. In appendix \ref{cond:1}, it is given that \eqref{cond:1a} and \eqref{cond:1b} are not problematic for this case also but we check that \eqref{cond:3} does not hold in appendix \ref{appendixcond:3}.
\subsection{Evaluation of \texorpdfstring{$\mathcal{A}_1$}{A1}}
We evaluate $\mathcal{T}_\mathrm{I}$.
\begin{align}
    \mathcal{T}_\mathrm{I}=\lim_{(\Lambda,\delta)\to(\infty,0)}\int_0^\infty\dd{L'}\int_0^\infty\dd{L}\ev{\psi(L+L')\mathcal{C}_\mathrm{I}(L,\Lambda,\delta)\psi(L)}_{C_{L+L'}}
\end{align}
In this case, because not only $\mathcal{C}_\mathrm{I}$ but also $\psi$ involves $X^0$, the correlation function cannot be factorized as \eqref{C1}. However we can derive
\begin{align}
    \ev{\partial X^0(x+iy)\bar{\partial}X^0(x-iy)e^{ik_1\cdot X(z_1)}\dots e^{ik_n\cdot X(z_n)}}_{C_{L+L'}}\propto e^{-2\pi y/(L+L')}
\end{align}
in the limit $y\to\infty$. Thus no matter what the matter operators which are involved in $\psi$, the horizontal part of the contours $P_{L,\Lambda,\delta},\bar{P}_{L,\Lambda,\delta}$ does not contribute $\mathcal{T}_\mathrm{I}$ in the limit $\Lambda\to\infty$. This leads to the same result as the one obtained in the previous section.
\begin{align}
    \mathcal{T}_\mathrm{I}&=\lim_{(\Lambda,\delta)\to(\infty,0)}\left[\ev{\psi(L+L')\int_{i\delta}^{i\Lambda}\frac{\dd{z}}{2\pi i}4\partial X^0(z)\bar{c}\bar{\partial}X^0(\bar{z})\psi(L)}_{C_{L+L'}}\right.\\
    &\qquad\qquad\qquad+\ev{\psi(L+L')\psi(L)\int_{i\delta}^{i\Lambda}\frac{\dd{z}}{2\pi i}4\partial X^0(z)\bar{c}\bar{\partial}X^0(\bar{z})}_{C_{L+L'}}\\
    &\qquad\qquad\qquad+\ev{\psi(L+L')\int_{-i\delta}^{-i\Lambda}\frac{\dd{\bar{z}}}{2\pi i}4\bar{\partial}X^0(\bar{z})\partial X^0(z)\psi(L)}_{C_{L+L'}}\\
    &\qquad\qquad\qquad\left.+\ev{\psi(L+L')\psi(L)\int_{-i\delta}^{-i\Lambda}\frac{\dd{\bar{z}}}{2\pi i}4\bar{\partial}X^0(\bar{z})\partial X^0(z)}_{C_{L+L'}}\right].
\end{align}

Next, we evaluate $\mathcal{T}_{\mathrm{I}\hspace{-1.2pt}\mathrm{I}\mathrm{A}}$. Because of the discussion in the previous section, we derive
\begin{align}
    &\ev{\psi(L+L')\mathcal{C}_{\mathrm{I}\hspace{-1.2pt}\mathrm{I}\mathrm{A}}(L,\Lambda,\delta)\psi(L)}_{C_{L+L'}}\\
    &=-2\left\langle \psi(L+L')\qty(c\partial X^0(i\infty)+\bar{c}\bar{\partial}X^0(-i\infty))\right.\\
    &\qquad\times\left.\frac{1}{2\pi i}\qty((X^0(L+i\delta)-X^0(L-i\delta))-(X^0(i\delta)-X^0(-i\delta)))\psi(L)\right\rangle_{C_{L+L'}}.
\end{align}
 In the limit $\delta\to0$, to avoid collision between $X^0$ and matter operators involved $\psi$, we regularize $\sqrt{F_{ji}}$ by
\begin{align}
    \sqrt{F_{ji}}=\lim_{\epsilon\to0}\int_\epsilon^\infty\dd{L}e^{-LK}f_{ji}(L).
\end{align}
Owing to the regularization, using the boundary condition \eqref{b.c.:X}, we obtain
\begin{align}
    \mathcal{T}_{\mathrm{I}\hspace{-1.2pt}\mathrm{I}\mathrm{A}}=0.
\end{align}

Finally, we evaluate $\mathcal{T}_{\mathrm{I}\hspace{-1.2pt}\mathrm{I}\mathrm{B}}$. In the same way as in the previous section, we need to consider only the anticommutator between $B$ and $\kappa$ because the contours $P_{L,\Lambda,\delta}$ and $\bar{P}_{L,\Lambda,\delta}$ do not cross the matter operators. Using \eqref{formula:kappafactorize}, we can derive
\begin{align}
    \lim_{\Lambda\to\infty}\ev{\psi(L+L')\acomm{\psi(L)}{\kappa(z,\bar{z})}}_{C_{L+L'}}=\frac{1}{L+1}\ev{\psi(L+L')\alpha(L)}_{C_{L+L'}},
\end{align}
where $\alpha$ is defined by
\begin{align}
    \alpha(L)\coloneqq\sum_i\alpha_i(L)\qc\alpha_i(L)\coloneqq\sum_i\mathcal{L}^{-1}\qty{\sqrt{F_{1i}}c\sqrt{F_{2i}}G_{1i}\frac{1}{H_i}G_{2i}\sqrt{F_{3i}}c\sqrt{F_{4i}}}
\end{align}
Thus as in the previous section, it is enough to consider
\begin{align}
    &\lim_{(\Lambda,\delta)\to(\infty,0)}\ev{\psi(L+L')\mathcal{C}_{\mathrm{I}\hspace{-1.2pt}\mathrm{I}\mathrm{B}}(L,\Lambda,\delta)\psi(L)}_{C_{L+L'}}\\
    &=\lim_{\delta\to0}\ev{\psi(L+L')\qty(\kappa(L+i\delta,L-i\delta)\psi(L)-\psi(L)\kappa(i\delta,-i\delta))}_{C_{L+L'}}\\
    &\qquad+\frac{1}{L+1}\ev{\psi(L+L')\alpha(L)}_{C_{L+L'}}.
\end{align}
Here using \eqref{formula:deltafactorize} we can derive
\begin{align}
    &\lim_{\delta\to0}\ev{\qty(c(i\delta)g_z(i\delta,-i\delta)-\bar{c}(-i\delta)g_{\bar{z}}(i\delta,-i\delta))e^{ik_1\cdot X(z_1)}\dots e^{ik_n\cdot X(z_n)}}_{C_{L+L'}}\\
    &=\lim_{\delta\to0}\left(\ev{g_z(i\delta,-i\delta)}_{C_{L+L'}}\ev{c(i\delta)e^{ik_1\cdot X(z_1)}\dots e^{ik_n\cdot X(z_n)}}_{C_{L+L'}}\right.\\
    &\qquad\qquad\left.-\ev{g_{\bar{z}}(i\delta,-i\delta)}_{C_{L+L'}}\ev{\bar{c}(-i\delta)e^{ik_1\cdot X(z_1)}\dots e^{ik_n\cdot X(z_n)}}_{C_{L+L'}}\right).
\end{align}
Hence even if $\psi$ involves matter operators, we can use
\begin{align}
    &\lim_{\delta\to0}\ev{\psi(L+L')\qty(\kappa(L+i\delta,L-i\delta)\psi(L)-\psi(L)\kappa(i\delta,-i\delta))}_{C_{L+L'}}\\
    &=\ev{\psi(L+L')\lim_{\delta\to0}\frac{1}{2\pi\delta}\qty(\qty(c(L+i\delta)-\bar{c}(L-i\delta))\psi(L)-\psi(L)\qty(c(i\delta)-\bar{c}(-i\delta)))}_{C_{L+L'}}.
\end{align}

Thus we obtain
\begin{align}
    &\lim_{(\Lambda,\delta)\to(\infty,0)}\ev{\psi(L+L')\mathcal{C}_{\mathrm{I}\hspace{-1.2pt}\mathrm{I}\mathrm{B}}(L,\Lambda,\delta)\psi(L)}_{C_{L+L'}}\\
    &=\lim_{\delta\to0}\ev{\psi(L+L')\frac{1}{2\pi\delta}\qty(\qty(c(L+i\delta)-\bar{c}(L-i\delta))\psi(L)-\psi(L)\qty(c(i\delta)-\bar{c}(-i\delta)))}_{C_{L+L'}}\\
    &\qquad+\frac{1}{L+1}\ev{\psi(L+L')\alpha(L)}_{C_{L+L'}}.
\end{align}
This is the same result as in the previous section. 

Therefore we obtain
\begin{align}
    \mathcal{A}_1=\int_0^\infty\dd{L'}\int_0^\infty\dd{L}\Tr(e^{-L'K}\psi(L')\qty(\chi e^{-LK}\psi(L)+e^{-LK}\psi(L)\chi+\frac{1}{1+L}e^{-LK}\alpha(L))).
\end{align}
\subsection{Evaluation of \texorpdfstring{$\mathcal{A}_2$}{A2}\label{result}}
We evaluate $\mathcal{A}_2$. In a similar way as in the previous section, we use
\begin{align}
    \ev{\psi(L+L')\mathcal{G}(L,\Lambda,\delta)\qty(Q\psi(L)-\mathcal{L}^{-1}\qty{Q\Psi}(L))}_{C_{L+L'}}=\eval{\partial_t\ev{\psi(L+L')\mathcal{G}(L,\Lambda,\delta)e^{-tK}\alpha(L+t)}_{C_{L+L'+t}}}_{t=0}.
\end{align}
Because $\psi$ involves $X^0$, this cannot be factorized as \eqref{factorize:A2}. If we focus on the case that $G_{1i}$ and $G_{2i}$ are constructed only by plane wave vertex operators, the right-hand side can be written by
\begin{align}
    &\eval{\partial_t\ev{\psi(L+L')\mathcal{G}(L,\Lambda,\delta)e^{-tK}\alpha(L+t)}_{C_{L+L'+t}}}_{t=0}\\
    &=\eval{\partial_t\qty(\ev{\mathcal{G}(L,\Lambda,\delta)}_{C_{L+L'+t}}\ev{\psi(L+L')e^{-tK}\alpha(L+t)}_{C_{L+L'+t}})}_{t=0}\\
    &\qquad+\sum_{i,n}\int\dd{L'_{1n}dL'_{2n}dL'_{3n}dL'_{4n}dL'_{5n}dL_{1i}dL_{2i}dL_{3i}dL_{4i}dL_{5i}}\\
    &\qquad\qquad\times f_{1n}(L'_{1n})f_{2n}(L'_{2n})h_n(L'_{3n})f_{3n}(L'_{4n})f_{4n}(L'_{5n})f_{1i}(L_{1i})f_{2i}(L_{2i})h_i(L_{3i})f_{3i}(L_{4i})f_{4i}(L_{5i})\\
    &\qquad\qquad\times\partial_t\left(\Delta(L_{1i}+L_{2i}+L_{3i}+L_{4i}+L_{5i},\Lambda,\delta,s+t)\right.\\
    &\qquad\qquad\qquad\times\Tr\left(e^{-L'_{1n}K}ce^{-L'_{2n}K}G_{1n}e^{-L'_{3n}K}BG_{2n}e^{-L'_{4n}K}ce^{-L'_{5n}K}\right.\\
    &\qquad\qquad\qquad\qquad\qquad\eval{\left.\left.e^{-tK}e^{-L_{1i}K}ce^{-L_{2i}K}G_{1i}e^{-L_{3i}K}G_{2i}e^{-L_{4i}K}ce^{-L_{5i}K}\right)\right)}_{t=0},
\end{align}
where $\Delta$ id defined by \eqref{def:Delta}. The first term can be written by
\begin{align}
    &\lim_{(\Lambda,\delta)\to(\infty,0)}\eval{\partial_t\qty(\ev{\mathcal{G}(L,\Lambda,\delta)}_{C_{L+L'+t}}\ev{\psi(L+L')e^{-tK}\alpha(L+t)}_{C_{L+L'+t}})}_{t=0}\\
    &=-\int_0^\infty\dd{L'}\int_0^\infty\dd{L}\frac{1}{1+L}\Tr(e^{-L'K}\psi(L')e^{-LK}\alpha(L))
\end{align}
in the same way as in the previous section. However the second term presents an obstruction. Because of the second term, we obtain
\begin{align}
    &\Tr(\Psi\comm{Q}{\mathcal{G}}\Psi)\\
    &=\Tr(\Psi(\chi-\chi^\dag)\Psi)\\
    &\qquad+\sum_{i,n}\int\dd{L'_{1n}dL'_{2n}dL'_{3n}dL'_{4n}dL'_{5n}dL_{1i}dL_{2i}dL_{3i}dL_{4i}dL_{5i}}\\
    &\qquad\qquad\times f_{1n}(L'_{1n})f_{2n}(L'_{2n})h_n(L'_{3n})f_{3n}(L'_{4n})f_{4n}(L'_{5n})f_{1i}(L_{1i})f_{2i}(L_{2i})h_i(L_{3i})f_{3i}(L_{4i})f_{4i}(L_{5i})\\
    &\qquad\qquad\times\lim_{(\Lambda,\delta)\to(\infty,0)}\partial_t\left(\Delta(L_{1i}+L_{2i}+L_{3i}+L_{4i}+L_{5i},\Lambda,\delta,s+t)\right.\\
    &\qquad\qquad\qquad\times\Tr\left(e^{-L'_{1n}K}ce^{-L'_{2n}K}G_{1n}e^{-L'_{3n}K}BG_{2n}e^{-L'_{4n}K}ce^{-L'_{5n}K}\right.\\
    &\qquad\qquad\qquad\qquad\qquad\eval{\left.\left.e^{-tK}e^{-L_{1i}K}ce^{-L_{2i}K}G_{1i}e^{-L_{3i}K}G_{2i}e^{-L_{4i}K}ce^{-L_{5i}K}\right)\right)}_{t=0}.
\end{align}
Here the trace in the last line leads to
\begin{align}
    &\Tr(e^{-L'_{1n}K}ce^{-L'_{2n}K}G_{1n}e^{-L'_{3n}K}BG_{2n}e^{-L'_{4n}K}ce^{-L'_{5n}K}e^{-tK}e^{-L_{1i}K}ce^{-L_{2i}K}G_{1i}e^{-L_{3i}K}G_{2i}e^{-L_{4i}K}ce^{-L_{5i}K})\\
    &=\frac{(s+t)^2}{4\pi^3}\left((L'_{1n}+L'_{5n}+t+L_{1i}+L_{2i}+L_{3i}+L_{4i}+L_{5i})\sin\frac{2\pi(L_{2i}+L_{3i}+L_{4i})}{s+t}\right.\\
    &\qquad+(L_{2i}+L_{3i}+L_{4i})\sin\frac{2\pi(L'_{1n}+L'_{5n}+t+L_{1i}+L_{2i}+L_{3i}+L_{4i}+L_{5i})}{s+t}\\
    &\qquad-(L'_{5n}+t+L_{1i}+L_{2i}+L_{3i}+L_{4i})\sin\frac{2\pi(L'_{1n}+L_{2i}+L_{3i}+L_{4i}+L_{5i})}{s+t}\\
    &\qquad-(L'_{1n}+L_{2i}+L_{3i}+L_{4i}+L_{5i})\sin\frac{2\pi(L'_{5n}+t+L_{1i}+L_{2i}+L_{3i}+L_{4i})}{s+t}\\
    &\qquad+(L'_{5n}+t+L_{1i})\sin\frac{2\pi(L'_{1n}+L_{5i})}{s+t}\\
    &\qquad\left.+(L'_{1n}+L_{5i})\sin\frac{2\pi(L'_{5n}+t+L_{1i})}{s+t}\right)\\
    &\times\Tr(e^{-(L'_{1n}+L'_{2n})K}G_{1n}e^{-L'_{3n}K}G_{2n}e^{-(L'_{4n}+L'_{5n}+t+L_{1i}+L_{2i})K}G_{1i}e^{-L_{3i}K}G_{2i}e^{-(L_{4i}+L_{5i})K}).
\end{align}
On the other hand, we were unable to evaluate $\Delta$. Hence it is not clear whether \eqref{cond:2} does not hold for the solution. If $\Delta$ does not vanishes, the difference between the energy and the Ellwood invariant is given by
\begin{align}
    &-S[\Psi_\mathrm{sol}]-\Tr(\mathcal{V}\Psi_\mathrm{sol})\\
    &=\sum_{i,n}\int\dd{L'_{1n}dL'_{2n}dL'_{3n}dL'_{4n}dL'_{5n}dL_{1i}dL_{2i}dL_{3i}dL_{4i}dL_{5i}}\\
    &\qquad\times f_{1n}(L'_{1n})f_{2n}(L'_{2n})h_n(L'_{3n})f_{3n}(L'_{4n})f_{4n}(L'_{5n})f_{1i}(L_{1i})f_{2i}(L_{2i})h_i(L_{3i})f_{3i}(L_{4i})f_{4i}(L_{5i})\\
    &\qquad\times\lim_{(\Lambda,\delta)\to(\infty,0)}\partial_t\left(\Delta(L_{1i}+L_{2i}+L_{3i}+L_{4i}+L_{5i},\Lambda,\delta,s+t)\right.\\
    &\qquad\qquad\times\frac{(s+t)^2}{4\pi^3}\left((L'_{1n}+L'_{5n}+t+L_{1i}+L_{2i}+L_{3i}+L_{4i}+L_{5i})\sin\frac{2\pi(L_{2i}+L_{3i}+L_{4i})}{s+t}\right.\\
    &\qquad+(L_{2i}+L_{3i}+L_{4i})\sin\frac{2\pi(L'_{1n}+L'_{5n}+t+L_{1i}+L_{2i}+L_{3i}+L_{4i}+L_{5i})}{s+t}\\
    &\qquad-(L'_{5n}+t+L_{1i}+L_{2i}+L_{3i}+L_{4i})\sin\frac{2\pi(L'_{1n}+L_{2i}+L_{3i}+L_{4i}+L_{5i})}{s+t}\\
    &\qquad-(L'_{1n}+L_{2i}+L_{3i}+L_{4i}+L_{5i})\sin\frac{2\pi(L'_{5n}+t+L_{1i}+L_{2i}+L_{3i}+L_{4i})}{s+t}\\
    &\qquad+(L'_{5n}+t+L_{1i})\sin\frac{2\pi(L'_{1n}+L_{5i})}{s+t}\\
    &\qquad\qquad\eval{\left.+(L'_{1n}+L_{5i})\sin\frac{2\pi(L'_{5n}+t+L_{1i})}{s+t}\right)}_{t=0}\\
    &\qquad\left.\times\Tr(e^{-(L'_{1n}+L'_{2n})K}G_{1n}e^{-L'_{3n}K}G_{2n}e^{-(L'_{4n}+L'_{5n}+t+L_{1i}+L_{2i})K}G_{1i}e^{-L_{3i}K}G_{2i}e^{-(L_{4i}+L_{5i})K})\right).
\end{align}
\section{Summary\label{summary}}
We examine condition \eqref{cond:2} for the solution which is constructed by $K,B,c$ and matter operators involving $X^0$. As a result, we obtain that $X^0$ presents an obstruction. If the solution involves $X^0$, we need to calculate $\Delta$. Hence \eqref{cond:2} may not hold for the solution. Because we confirm that \eqref{cond:1a} and \eqref{cond:1b} are not problematic in appendix \ref{cond:1}, if we can evaluate $\Delta$, it will be clear whether the energy is proportional to the Ellwood invariant. Unfortunately, $\Delta$ depends on $X^0$ included in the solution and we were unable to evaluate $\Delta$. Thus at present, it is not clear whether the energy is proportional to the Ellwood invariant. However, according to the numerical result in Appendix \ref{Ape:Xcorr}, $\Delta$ does not vanish (Figure \ref{fig:hatDeltaintegral}). Therefore the energy may be not proportional to the Ellwood invariant.

If one would like to clarify whether the energy is proportional to the Ellwood invariant for such solutions, it may solve the problem to modify $\mathcal{G}$. It is required that $\mathcal{G}$ satisfies
\begin{align}
    \comm{Q}{\mathcal{G}}=\chi-\chi^\dag,
\end{align}
but such $\mathcal{G}$ is not unique. In \cite{Hata:2021lqz,Hata:2022kfq}, they found operator sets that satisfy the algebraic relation of the $KBc$ algebra. Using such operator sets even if a solution involves $X^0$, it looks like the $KBc$ solution. They may be helpful to modify $\mathcal{G}$.

In this paper, we focused on regular solutions and did not consider solutions in which regularization is necessary e.g. \cite{Murata:2011ep,Murata:2011ex,Hata:2019dwu,Hata:2019ybw}. In \cite{Baba:2012cs}, it is already examined for Murata-Schnabl solution but it may be interesting to examine also for other solutions. Especially the solution which is constructed in \cite{Miwa:2017oxy} involves $X^0$ and regularization is necessary. It would be intriguing to examine the relation between the energy and the Ellwood invariant for the solution.
\section*{Acknowledgements}
The authors would like to thank Nobuyuki Ishibashi for reading a preliminary draft and helpful comments. Additionally, the authors would like to thank a referee for proposing a numerical approach to \eqref{Deltaafterlimit} in Appendix \ref{Ape:Xcorr}.
The work of Y.A. was supported by JST, the establishment of university fellowships towards the creation of science technology innovation, Grant Number JPMJFS2106.

\appendix
\section{Correlation functions in the sliver frame\label{Ape:Xcorr}}
The conformal transformation $f_s$ from the infinite cylinder $C_s$ with circumference $s$ to upper half plane (UHP) and the inverse transformation $f_s^{-1}$ are given as
\begin{alignat}{2}
    f_s(u)&=\frac{s}{\pi}\tan^{-1}u&\qc f_s^{-1}(z)&=\tan\frac{\pi}{s}z,\label{def:slivermap}\\
    \dv{f_s}{u}&=\frac{s}{\pi}\frac{1}{1+u^2}&\qc\dv{f_s^{-1}}{z}&=\frac{\pi}{s}\frac{1}{\cos^2\frac{\pi}{s}z}.
\end{alignat}
Using them, we obtain the 2-point correlation function of $\partial X$ on $C_s$.
\begin{align}
    \ev{\partial X^\mu(z_1)\partial X^\nu(z_2)}_{C_s}&=\ev{f_s^{-1}\circ(\partial X^\mu(z_1)\partial X^\nu(z_2))}_\mathrm{UHP}\\
    &=-\frac{1}{2}\eta^{\mu\nu}\qty(\frac{\pi}{s})^2\frac{1}{\sin^2\frac{\pi(z_1-z_2)}{s}}
\end{align}
Especially, since for
\begin{align}
    X^0(z,\bar{z})-X^0(z_0,\bar{z}_0)=\int_{z_0}^z\dd{z'}\partial X^0(z')+\int_{\bar{z}_0}^{\bar{z}}\dd{\bar{z}'}\bar{\partial}X^0(\bar{z}'),
\end{align}
we can evaluate the correlation function of $g_z$
\begin{align}
    \ev{g_z(z,\bar{z})}_{C_s}&=\lim_{z_0\to i\infty}2\ev{\qty(X^0(z,\bar{z})-X^0(z_0,\bar{z}_0))\partial X^0(z}_{C_s}\\
    &=\lim_{z_0\to i\infty}2\qty(\int_{z_0}^z\ev{\partial X^0(z')\partial X^0(z)}_{C_s}+\int_{\bar{z}_0}^{\bar{z}}\dd{\bar{z}'}\ev{\bar{\partial}X^0(\bar{z}')\partial X^0(z)}_{C_s})\\
    &=\frac{\pi}{s}\cot\frac{\pi(z-\bar{z})}{s}\label{correlation:g}
\end{align}
and the correlation function of $\mathcal{G}$
\begin{align}
    \ev{\mathcal{G}(L,\Lambda,\delta)}_{C_s}=\frac{L}{s}\coth\frac{2\pi\Lambda}{s}.\label{correlation:G}
\end{align}

Similarly, we obtain the correlation function of vertex operators on $C_s$
\begin{align}
    \ev{e^{ik_1\cdot X(z_1)}\dots e^{ik_n\cdot X(z_n)}}_{C_s}&=\prod_{i=1}^n\qty(\frac{s}{\pi}\cos^2\frac{\pi}{s}z_i)^{-k_i^2/2}\ev{e^{ik_1\cdot X(u_1)}\dots e^{ik_n\cdot X(u_n)}}_\mathrm{UHP}\\
    &=\prod_{i=1}^n\qty(\frac{s}{\pi}\cos^2\frac{\pi}{s}z_i)^{-k_i^2/2}\prod_{1\le j<l\le n}\qty(\tan\frac{\pi}{s}z_j-\tan\frac{\pi}{s}z_l)^{k_j\cdot k_l}\\
    &=\prod_{i=1}^n\qty(\frac{s}{\pi}\cos^2\frac{\pi}{s}z_i)^{-k_i^2/2}\prod_{1\le j<l\le n}\qty(\frac{\sin\frac{\pi}{s}(z_j-z_l)}{\cos\frac{\pi}{s}z_j\cos\frac{\pi}{s}z_l})^{k_j\cdot k_l}\\
    &=\exp[-\sum_{i=1}^n\frac{k_i^2}{2}\ln(\frac{s}{\pi}\cos^2\frac{\pi}{s}z_i)+\sum_{1\le j<l\le n}k_j\cdot k_l\ln(\frac{\sin\frac{\pi}{s}(z_j-z_l)}{\cos\frac{\pi}{s}z_j\cos\frac{\pi}{s}z_l})],
\end{align}
where $u_i$ is defined by
\begin{align}
    u_i\coloneqq\tan\frac{\pi}{s}z_i.
\end{align}
In general, correlation functions of $X$ variables can be evaluated from above result. For example, the correlation function of $\partial X^\mu e^{ik\cdot X}$ is given by
\begin{align}
    \ev{X^\mu(z')e^{ik\cdot X}(z)}_{C_s}&=\eval{\frac{1}{i}\pdv{k'_\mu}\ev{e^{ik'\cdot X(z')}e^{ik\cdot X(z)}}_{C_s}}_{k'=0}\\
    &=-ik^\mu\ln(\frac{\sin\frac{\pi}{s}(z'-z)}{\cos\frac{\pi}{s}z'\cos\frac{\pi}{s}z}),\\
    \ev{\partial X^\mu(z')e^{ik\cdot X}(z)}_{C_s}&=\pdv{z'}\ev{X^\mu(z')e^{ik\cdot X}(z)}_{C_s}\\
    &=-ik^\mu\frac{\pi}{s}\qty(\tan\frac{\pi}{s}z'+\cot\frac{\pi}{s}(z'-z)).
\end{align}
In particular, we give important correlation function.
\begin{align}
    &\ev{\partial X^\mu(z')\partial X^\nu(z)e^{ik_1\cdot X(z_1)}\dots e^{ik_n\cdot X(z_n)}}_{C_s}\\
    &=\qty[\ev{\partial X^\mu(z')\partial X^\nu(z)}_{C_s}+\sum_{\substack{1\le m\le n\\1\le o\le n}}\ev{\partial X^\mu(z')e^{ik_m\cdot X}(z_m)}_{C_s}\ev{\partial X^\nu(z)e^{ik_o\cdot X}(z_o)}]\ev{e^{ik_1\cdot X(z_1)}\dots e^{ik_n\cdot X(z_n)}}_{C_s}\\
    &=\qty[\ev{\partial X^\mu(z')\partial X^\nu(z)}_{C_s}-\qty(\frac{\pi}{s})^2\sum_{\substack{1\le m\le n\\1\le o\le n}}(k_m)^\mu(k_o)^\nu\qty(\tan\frac{\pi}{s}z'+\cot\frac{\pi}{s}(z'-z_m))\qty(\tan\frac{\pi}{s}z+\cot\frac{\pi}{s}(z-z_o))]\\
    &\qquad\times\ev{e^{ik_1\cdot X(z_1)}\dots e^{ik_n\cdot X(z_n)}}_{C_s}.
\end{align}

Using them, we obtain
\begin{align}
    &\ev{(X^\mu(z,\bar{z})-X^\mu(z_0,\bar{z}_0))\partial X^\nu(z)e^{ik_1\cdot X(z_1)}\dots e^{ik_n\cdot X(z_n)}}_{C_s}\\
    &=\ev{(X^\mu(z,\bar{z})-X^\mu(z_0,\bar{z}_0))\partial X^\nu(z)}_{C_s}\ev{e^{ik_1\cdot X(z_1)}\dots e^{ik_n\cdot X(z_n)}}_{C_s}\\
    &\qquad-\frac{\pi}{s}\sum_{\substack{1\le m\le n\\1\le o\le n}}(k_m)^\mu(k_o)^\nu\ev{e^{ik_1\cdot X(z_1)}\dots e^{ik_n\cdot X(z_n)}}_{C_s}\qty(\tan\frac{\pi}{s}z+\cot\frac{\pi}{s}(z-z_o))\\
    &\qquad\qquad\times\ln(\frac{-\cos\frac{2\pi}{s}(\Re z-z_m)+\cos\frac{2\pi i}{s}\Im z}{\cos\frac{2\pi}{s}\Re z+\cos\frac{2\pi i}{s}\Im z}\frac{\cos\frac{2\pi}{s}\Re z_0+\cos\frac{2\pi i}{s}\Im z_0}{-\cos\frac{2\pi}{s}(\Re z_0-z_m)+\cos\frac{2\pi i}{s}\Im z_0}).
\end{align}
Since for
\begin{align}
    \ln(\frac{\cos\frac{2\pi}{s}\Re z_0+\cos\frac{2\pi i}{s}\Im z_0}{-\cos\frac{2\pi}{s}(\Re z_0-z_m)+\cos\frac{2\pi i}{s}\Im z_0})=4\cos\frac{\pi}{s}z_m\cos(\frac{\pi}{s}(2x_0-z_m))e^{2\pi ix_0}e^{-2\pi y_0/s}+\order{e^{-4\pi y_0/s}},
\end{align}
with $z_0=x_0+iy_0$ in the limit $y_0\to\infty$, we obtain
\begin{align}
    &\ev{(X^\mu(z,\bar{z})-X^\mu(i\infty,-i\infty))\partial X^\nu(z)e^{ik_1\cdot X(z_1)}\dots e^{ik_n\cdot X(z_n)}}_{C_s}\\
    &=\ev{(X^\mu(z,\bar{z})-X^\mu(i\infty,-i\infty))\partial X^\nu(z)}_{C_s}\ev{e^{ik_1\cdot X(z_1)}\dots e^{ik_n\cdot X(z_n)}}_{C_s}\\
    &\qquad-\frac{\pi}{s}\sum_{\substack{1\le m\le n\\1\le o\le n}}(k_m)^\mu(k_o)^\nu\qty(\tan\frac{\pi}{s}z+\cot\frac{\pi}{s}(z-z_o))\ln(\frac{-\cos\frac{2\pi}{s}(\Re z-z_m)+\cos\frac{2\pi i}{s}\Im z}{\cos\frac{2\pi}{s}\Re z+\cos\frac{2\pi i}{s}\Im z})\\
    &\qquad\qquad\times\ev{e^{ik_1\cdot X(z_1)}\dots e^{ik_n\cdot X(z_n)}}_{C_s},
\end{align}
and
\begin{align}
    \ev{g_z(z,\bar{z})e^{ik_1\cdot X(z_1)}\dots e^{ik_n\cdot X(z_n)}}_{C_s}&=\qty(\ev{g_z(z,\bar{z})}_{C_s}+\delta_z(z,\bar{z},s))\ev{e^{ik_1\cdot X(z_1)}\dots e^{ik_n\cdot X(z_n)}}_{C_s},\\    
    \ev{g_{\bar{z}}(z,\bar{z})e^{ik_1\cdot X(z_1)}\dots e^{ik_n\cdot X(z_n)}}_{C_s}&=\qty(\ev{g_{\bar{z}}(z,\bar{z})}_{C_s}+\delta_{\bar{z}}(z,\bar{z},s))\ev{e^{ik_1\cdot X(z_1)}\dots e^{ik_n\cdot X(z_n)}}_{C_s},
\end{align}
where we define
\begin{align}
    \begin{alignedat}{1}
        \delta_z(z,\bar{z},s)&\coloneqq\sum_{\substack{1\le m\le n\\1\le o\le n}}\ev{(X^0(z,\bar{z})-X^0(i\infty,-i\infty))e^{ik_m\cdot X(z_m)}}_{C_s}\ev{\partial X^0(z)e^{ik_o\cdot X(z_o)}}_{C_s}\\
        &=-\frac{2\pi}{s}\sum_{\substack{1\le m\le n\\1\le o\le n}}(k_m)^0(k_o)^0\qty(\tan\frac{\pi}{s}z+\cot\frac{\pi}{s}(z-z_o))\ln(\frac{-\cos\frac{2\pi}{s}(\Re z-z_m)+\cos\frac{2\pi}{s}\Im z}{\cos\frac{2\pi}{s}\Re z+\cos\frac{2\pi}{s}\Im z}),\\
        \delta_{\bar{z}}(z,\bar{z},s)&\coloneqq\sum_{\substack{1\le m\le n\\1\le o\le n}}\ev{(X^0(z,\bar{z})-X^0(i\infty,-i\infty))e^{ik_m\cdot X(z_m)}}_{C_s}\ev{\bar{\partial}X^0(\bar{z})e^{ik_o\cdot X(z_o)}}_{C_s}\\
        &=-\frac{2\pi}{s}\sum_{\substack{1\le m\le n\\1\le o\le n}}(k_m)^0(k_o)^0\qty(\tan\frac{\pi}{s}\bar{z}+\cot\frac{\pi}{s}(\bar{z}-z_o))\ln(\frac{-\cos\frac{2\pi}{s}(\Re z-z_m)+\cos\frac{2\pi}{s}\Im z}{\cos\frac{2\pi}{s}\Re z+\cos\frac{2\pi}{s}\Im z}).
    \end{alignedat}
\end{align}

Since for
\begin{align}
    \tan\frac{\pi}{s}(x+iy)+\cot\frac{\pi}{s}(x+iy-z_o)&=-4ie^{\pi i(2x-z_o)/s}\cos(\frac{\pi}{s}z_o)e^{-2\pi y/s}+\order{e^{-4\pi y/s}},\label{lim:partialXe}\\
    \ln(\frac{-\cos\frac{2\pi}{s}(\Re z-z_m)+\cos\frac{2\pi}{s}\Im z}{\cos\frac{2\pi}{s}\Re z+\cos\frac{2\pi}{s}\Im z})&=-4\cos\frac{\pi}{s}z_m\cos(\frac{\pi}{s}(2x-z_m))e^{2\pi ix}e^{-2\pi y/s}+\order{e^{-4\pi y/s}},
\end{align}
with $z=x+iy$ in the limit $y\to\infty$, we obtain
\begin{align}
    \lim_{\Lambda\to\infty}\delta_z(i\Lambda,-i\Lambda,s)=0\qc\lim_{\Lambda\to\infty}\delta_{\bar{z}}(i\Lambda,-i\Lambda,s)=0,
\end{align}
and
\begin{align}
    \begin{alignedat}{1}
        \lim_{\Lambda\to\infty}\ev{g_z(i\Lambda,-i\Lambda)e^{ik_1\cdot X(z_1)}\dots e^{ik_n\cdot X(z_n)}}_{C_s}&=\lim_{\Lambda\to\infty}\ev{g_z(i\Lambda,-i\Lambda)}_{C_s}\ev{e^{ik_1\cdot X(z_1)}\dots e^{ik_n\cdot X(z_n)}}_{C_s},\\
        \lim_{\Lambda\to\infty}\ev{g_{\bar{z}}(i\Lambda,-i\Lambda)e^{ik_1\cdot X(z_1)}\dots e^{ik_n\cdot X(z_n)}}_{C_s}&=\lim_{\Lambda\to\infty}\ev{g_{\bar{z}}(i\Lambda,-i\Lambda)}_{C_s}\ev{e^{ik_1\cdot X(z_1)}\dots e^{ik_n\cdot X(z_n)}}_{C_s}.
    \end{alignedat}\label{formula:deltafactorize}
\end{align}

Using the above results, we give the correlation function of $\mathcal{G}$ and plane wave vertex operators
\begin{align}
    \ev{\mathcal{G}(L,\Lambda,\delta)e^{ik_1\cdot X(z_1)}\dots e^{ik_n\cdot X(z_n)}}_{C_s}&=\qty(\ev{\mathcal{G}(L,\Lambda,\delta)}_{C_s}+\Delta(L,\Lambda,\delta,s))\ev{e^{ik_1\cdot X(z_1)}\dots e^{ik_n\cdot X(z_n)}}_{C_s},\\
    \Delta(L,\Lambda,\delta,s)&\coloneqq\int_{P_{L,\Lambda,\delta}}\frac{\dd{z}}{2\pi i}\delta_z(z,\bar{z},s)-\int_{\bar{P}_{L,\Lambda,\delta}}\frac{\dd{\bar{z}}}{2\pi i}\delta_{\bar{z}}(z,\bar{z},s).\label{def:Delta}
\end{align}
In the limit $\Lambda\to\infty$, the horizontal part of the integration vanishes
\begin{align}
    \lim_{\Lambda\to\infty}\int_{i\Lambda}^{L+i\Lambda}\frac{\dd{z}}{2\pi i}\delta_z(z,\bar{z},s)=0\qc\lim_{\Lambda\to\infty}\int_{-i\Lambda}^{L-i\Lambda}\frac{\dd{\bar{z}}}{2\pi i}\delta_{\bar{z}}(z,\bar{z},s)=0.
\end{align}
On the other hand, the vertical part of the integration does not vanish. Thus in the limit $\Lambda\to\infty$, we obtain
\begin{align}
    \lim_{\Lambda\to\infty}\Delta(L,\Lambda,\delta,s)&=\lim_{\Lambda\to\infty}\left[\int_{i\delta}^{i\Lambda}\frac{\dd{z}}{2\pi i}\delta_z(z,\bar{z},s)-\int_{-i\delta}^{-i\Lambda}\frac{\dd{\bar{z}}}{2\pi i}\delta_{\bar{z}}(z,\bar{z},s)\right.\\
    &\qquad\left.\qquad-\int_{L+i\delta}^{L+i\Lambda}\frac{\dd{z}}{2\pi i}\delta_z(z,\bar{z},s)+\int_{L-i\delta}^{L-i\Lambda}\frac{\dd{\bar{z}}}{2\pi i}\delta_{\bar{z}}(z,\bar{z},s)\right].\label{Deltaafterlimit}
\end{align}
Unfortunately, we are unable to evaluate this integral and we do not know whether this vanishes. 

\begin{align}
    \lim_{\Lambda\to\infty}\Delta(L,\Lambda,\delta,s)&=\lim_{\Lambda\to\infty}\left[\int_{i\delta}^{i\Lambda}\frac{\dd{z}}{2\pi i}\delta_z(z,\bar{z},s)-\int_{-i\delta}^{-i\Lambda}\frac{\dd{\bar{z}}}{2\pi i}\delta_z(\bar{z},z,s)\right.\\
    &\qquad\left.\qquad-\int_{L+i\delta}^{L+i\Lambda}\frac{\dd{z}}{2\pi i}\delta_z(z,\bar{z},s)+\int_{L-i\delta}^{L-i\Lambda}\frac{\dd{\bar{z}}}{2\pi i}\delta_{\bar{z}}(z,\bar{z},s)\right].
\end{align}

\begin{figure}
    \centering
    \includegraphics{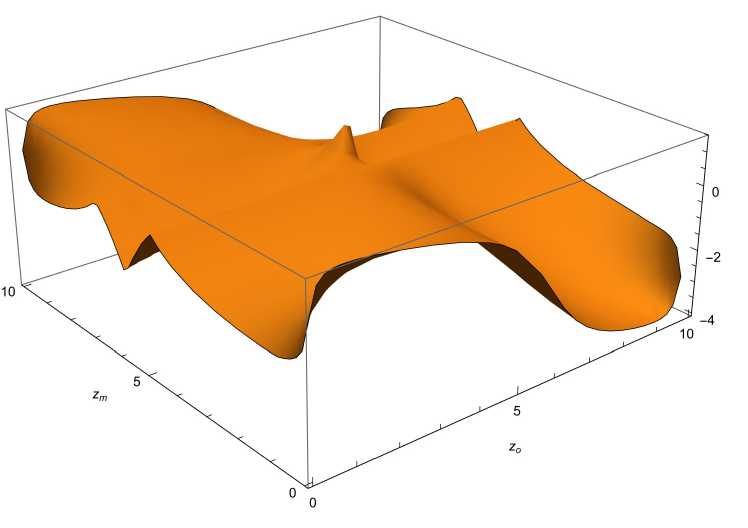}
    \caption{The numerical result for $\Im\hat{\Delta}(10,6,z_m,z_o)$}
    \label{fig:hatDeltaintegral}
\end{figure}

We numerically evaluate
\begin{align}
    \hat{\Delta}(s,L,z_m,z_o)&\coloneqq\int_0^{i\infty}\frac{\dd{z}}{2\pi i}\hat{\delta}_z(z,\bar{z},s,z_m,z_o)-\int_0^{-i\infty}\frac{\dd{z}}{2\pi i}\hat{\delta}_z(z,\bar{z},s,z_m,z_o)\\
    &-\int_L^{L+i\infty}\frac{\dd{z}}{2\pi i}\hat{\delta}_z(z,\bar{z},s,z_m,z_o)+\int_L^{L-i\infty}\frac{\dd{z}}{2\pi i}\hat{\delta}_{\bar{z}}(z,\bar{z},s,z_m,z_o)\label{hatDeltaintegral}
\end{align}
and in Figure \ref{fig:hatDeltaintegral}, we plot $\Re\hat{\Delta}(10,6,z_m,z_o)$ as a function of $z_m,z_o$ where $\hat{\delta}_z,\hat{\delta}_{\bar{z}}$ are defined by
\begin{align}
    \hat{\delta}_z(z,\bar{z},s,z_m,z_o)&\coloneqq-\frac{2\pi}{s}\qty(\tan\frac{\pi}{s}z+\cot\frac{\pi}{s}(z-z_o))\ln(\frac{-\cos\frac{2\pi}{s}(\Re z-z_m)+\cos\frac{2\pi}{s}\Im z}{\cos\frac{2\pi}{s}\Re z+\cos\frac{2\pi}{s}\Im z}),\\
    \hat{\delta}_{\bar{z}}(z,\bar{z},s,z_m,z_o)&\coloneqq-\frac{2\pi}{s}\qty(\tan\frac{\pi}{s}z+\cot\frac{\pi}{s}(\bar{z}-z_o))\ln(\frac{-\cos\frac{2\pi}{s}(\Re z-z_m)+\cos\frac{2\pi}{s}\Im z}{\cos\frac{2\pi}{s}\Re z+\cos\frac{2\pi}{s}\Im z}).
\end{align}
According to Figure \ref{fig:hatDeltaintegral}, It is clear that $\Delta$ does not vanish in general.
\section{Examination of condition (\ref{cond:3})\label{appendixcond:3}}
Even if $\Delta$ does not vanish, it does not imply that the energy is absolutely not proportion to the Ellwood invariant. This is because there is still a possibility that \eqref{cond:3} does not hold.
\begin{align}
    \Tr(\frac{2}{\pi i}c\bar{c}\partial X^0\bar{\partial}X^0\Psi)\neq\Tr(\chi Q\Psi)
\end{align}
In the this appendix, we show that the above relation for string fields involving $X^0$ variables does not hold.

First let us consider a string field
\begin{align}
    \Psi_{L,L'}\coloneqq e^{-LK}ce^{-L'K}.
\end{align}
The Ellwood invariant for the string field is
\begin{align}
    \Tr(\frac{2}{\pi i}c\bar{c}\partial X^0\bar{\partial}X^0\Psi_{L,L'})&=\frac{2}{\pi i}\ev{c\bar{c}\partial X^0\bar{\partial}X^0(i\infty,-i\infty)c(-L)}_{C_s}\\
    &=\frac{2}{\pi i}\ev{c\bar{c}(i\infty,-i\infty)c(-L)}_{C_s}\ev{\partial X^0\bar{\partial}X^0(i\infty,-i\infty)}_{C_s}.
\end{align}
On the other hand, the trace of $\chi Q\Psi_{L,L'}$ is given by
\begin{align}
    &\Tr(\chi Q\Psi_{L,L'})\\
    &=\lim_{(\Lambda,\delta)\to(\infty,0)}\left[\int_{i\delta}^{i\Lambda}\frac{\dd{z}}{2\pi i}4\ev{\partial X^0(z)\bar{c}\bar{\partial}X^0(\bar{z})c\partial c(-L)}_{C_s}-\int_{-i\delta}^{-i\Lambda}\frac{\dd{\bar{z}}}{2\pi i}4\ev{\bar{\partial}X^0(\bar{z})c\partial X^0(z)c\partial c(-L)}_{C_s}\right.\\
    &\qquad\qquad\left.+\frac{1}{2\pi\delta}\ev{c(0)c\partial c(-L)}_{C_s}\right]\\
    &=\lim_{(\Lambda,\delta)\to(\infty,0)}\left[\int_{i\delta}^{i\Lambda}\frac{\dd{z}}{2\pi i}4\ev{\bar{c}(\bar{z})c\partial c(-L)}_{C_s}\ev{\partial X^0(z)\bar{\partial}X^0(\bar{z})}_{C_s}\right.\\
    &\qquad\qquad\left.-\int_{-i\delta}^{-i\Lambda}\frac{\dd{\bar{z}}}{2\pi i}4\ev{c(z)c\partial c(-L)}_{C_s}\ev{\bar{\partial}X^0(\bar{z})\partial X^0(z)}_{C_s}+\frac{1}{2\pi\delta}\ev{c(0)c\partial c(-L)}_{C_s}\right].
\end{align}
As shown in \cite{Baba:2012cs}, they are the same.
\begin{align}
    \Tr(\frac{2}{\pi i}c\bar{c}\partial X^0\bar{\partial}X^0\Psi_{L,L'})=\Tr(\chi Q\Psi_{L,L'})\label{relation:Ellwoodchi}
\end{align}

Next let us consider a string field
\begin{align}
    \Psi=e^{-L_1K}e^{ik_1\cdot X}e^{-L_2K}ce^{-L_3K}e^{ik_2\cdot X}e^{-L_4K},
\end{align}
where we set $k_1^\mu=(\sqrt{h},0,\dots,0)$ and $k_2^\mu=-k_1^\mu$. The Ellwood invariant for the string field is
\begin{align}
    &\Tr(\frac{2}{\pi i}c\bar{c}\partial X^0\bar{\partial}X^0\Psi)\\
    &=\frac{2}{\pi i}\ev{c\bar{c}\partial X^0\bar{\partial}X^0(i\infty,-i\infty)e^{ik_1\cdot X}(-L_1)c(-L_1-L_2)e^{ik_2\cdot X}(-L_1-L_2-L_3)}_{C_s}\\
    &=\frac{2}{\pi i}\ev{c\bar{c}(i\infty,-i\infty)c(-L_1-L_2)}_{C_s}\ev{\partial X^0\bar{\partial}X^0(i\infty,-i\infty)e^{ik_1\cdot X}(-L_1)e^{ik_2\cdot X}(-L_1-L_2-L_3)}_{C_s}\\
    &=\frac{2}{\pi i}\ev{c\bar{c}(i\infty,-i\infty)c(-L_1-L_2)}_{C_s}\ev{e^{ik_1\cdot X}(z_1)e^{ik_2\cdot X}(z_2)}_{C_s}\\
    &\qquad\times\qty(\ev{\partial X^0\bar{\partial}X^0(i\infty,-i\infty)}_{C_s}+\sum_{\substack{1\le m\le2\\1\le o\le2}}\ev{\partial X^0(i\infty)e^{ik_m\cdot X}(z_m)}_{C_s}\ev{\bar{\partial}X^0(-i\infty)e^{ik_o\cdot X}(z_o))}_{C_s}),
\end{align}
where
\begin{align}
    z_1=-L_1\qc z_2=-L_1-L_2-L_3.
\end{align}
Since for \eqref{lim:partialXe}, we obtain
\begin{align}
    \ev{\partial X^0(i\infty)e^{ik_m\cdot X}(z_m)}_{C_s}=\ev{\bar{\partial}X^0(-i\infty)e^{ik_o\cdot X}(z_o)}_{C_s}=0.
\end{align}
Thus the Ellwood invariant is
\begin{align}
    &\Tr(\frac{2}{\pi i}c\bar{c}\partial X^0\bar{\partial}X^0\Psi)\\
    &=\frac{2}{\pi i}\ev{c\bar{c}(i\infty,-i\infty)c(-L_1-L_2)}_{C_s}\ev{\partial X^0\bar{\partial}X^0(i\infty,-i\infty)}_{C_s}\ev{e^{ik_1\cdot X}(z_1)e^{ik_2\cdot X}(z_2)}_{C_s}\\
    &=\Tr(\frac{2}{\pi i}c\bar{c}\partial X^0\bar{\partial}X^0\Psi_{L_1+L_2,L_3+L_4})\ev{e^{ik_1\cdot X}(z_1)e^{ik_2\cdot X}(z_2)}_{C_s}\\
    &=\Tr'\qty(\frac{2}{\pi i}c\bar{c}\partial X^0\bar{\partial}X^0\Psi_{L_1+L_2,L_3+L_4}),\label{EllwoodX0}
\end{align}
where
\begin{align}
    \Tr'(\Psi\Phi)\coloneqq\Tr(\Psi\Phi)\ev{e^{ik_1\cdot X}(z_1)e^{ik_2\cdot X}(z_2)}_{C_s}.
\end{align}

\begin{figure}
    \centering
    \includegraphics{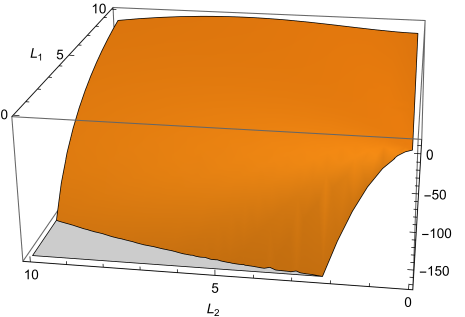}
    \caption{The numerical result for \eqref{result:chi} where we set $L_1=L_4,L_2=L_3$ and $h=1$.}
    \label{fig:chi}
\end{figure}

Finally, we evaluate the trace of $\chi Q\Psi$. $Q\Psi$ is 
\begin{align}
    Q\Psi&=e^{-L_1K}Qe^{ik_1\cdot X}e^{-L_2K}ce^{-L_3K}e^{ik_2\cdot X}e^{-L_4K}+e^{-L_1K}e^{ik_1\cdot X}e^{-L_2K}c\partial ce^{-L_3K}e^{ik_2\cdot X}e^{-L_4K}\\
    &\qquad-e^{-L_1K}e^{ik_1\cdot X}e^{-L_2K}ce^{-L_3K}Qe^{ik_2\cdot X}e^{-L_4K}.\label{testfieldBRS}
\end{align}
We focus on the second term and we consider the correlation function of it and the first term in \eqref{def:chi}. This is given by
\begin{align}
    &\lim_{(\Lambda,\delta)\to(\infty,0)}\int_{i\delta}^{i\Lambda}\frac{\dd{z}}{2\pi i}4\ev{\partial X^0(z)\bar{c}\bar{\partial}X^0(\bar{z})e^{ik_1\cdot X}(-L_1)c\partial c(-L_1-L_2)e^{ik_2\cdot X}(-L_1-L_2-L_3)}_{C_s}\\
    &=\lim_{(\Lambda,\delta)\to(\infty,0)}\int_{i\delta}^{i\Lambda}\frac{\dd{z}}{2\pi i}4\ev{\bar{c}(\bar{z})c\partial c(-L_1-L_2)}_{C_s}\ev{\partial X^0(z)\bar{\partial}X^0(\bar{z})e^{ik_1\cdot X}(-L_1)e^{ik_2\cdot X}(-L_1-L_2-L_3)}_{C_s}\\
    &=\lim_{(\Lambda,\delta)\to(\infty,0)}\int_{i\delta}^{i\Lambda}\frac{\dd{z}}{2\pi i}4\ev{\bar{c}(\bar{z})c\partial c(-L_1-L_2)}_{C_s}\\
    &\qquad\times\qty(\ev{\partial X^0(z)\bar{\partial}X^0(\bar{z})}_{C_s}+\sum_{\substack{1\le m\le2\\1\le o\le2}}\ev{\partial X^0(z)e^{ik_m\cdot X}(z_m)}_{C_s}\ev{\bar{\partial}X^0(\bar{z})e^{ik_o\cdot X}(z_o)}_{C_s})\\
    &\qquad\times\ev{e^{ik_1\cdot X}(-L_1)e^{ik_2\cdot X}(-L_1-L_2-L_3)}_{C_s}.
\end{align}
Similarly, we obtain
\begin{align}
    &\Tr'\qty(\chi e^{-L_1K}e^{ik_1\cdot X}e^{-L_2K}c\partial ce^{-L_3K}e^{ik_2\cdot X}e^{-L_4K})\\
    &=\Tr'\qty(\chi Q\Psi_{L_1+L_2,L_3+L_4})\\
    &+\lim_{(\Lambda,\delta)\to(\infty,0)}\sum_{\substack{1\le m\le2\\1\le o\le2}}\left(\int_{i\delta}^{i\Lambda}\frac{\dd{z}}{2\pi i}4\ev{\bar{c}(\bar{z})c\partial c(-L_1-L_2)}_{C_s}\ev{\partial X^0(z)e^{ik_m\cdot X}(z_m)}_{C_s}\ev{\bar{\partial}X^0(\bar{z})e^{ik_o\cdot X}(z_o)}_{C_s}\right.\\
    &\qquad\qquad\qquad\left.-\int_{-i\delta}^{-i\Lambda}\frac{\dd{\bar{z}}}{2\pi i}4\ev{c(z)c\partial c(-L_1-L_2)}_{C_s}\ev{\partial X^0(z)e^{ik_m\cdot X}(z_m)}_{C_s}\ev{\bar{\partial}X^0(\bar{z})e^{ik_o\cdot X}(z_o)}_{C_s}\right).\label{piece1}
\end{align}
We evaluate also the the trace of $\chi$ and the remaining term in \eqref{testfieldBRS}. They are given by
\begin{align}
    &\Tr(\chi e^{-L_1K}Qe^{ik_1\cdot X}e^{-L_2K}ce^{-L_3K}e^{ik_2\cdot X}e^{-L_4K})\\
    &=-h\Tr(\chi e^{-L_1K}\partial ce^{ik_1\cdot X}e^{-L_2K}ce^{-L_3K}e^{ik_2\cdot X}e^{-L_4K})+i\sqrt{h}\Tr(\chi e^{-L_1K}c\partial X^0e^{ik_1\cdot X}e^{-L_2K}ce^{-L_3K}e^{ik_2\cdot X}e^{-L_4K})\\
    &=-h\lim_{(\Lambda,\delta)\to(\infty,0)}\left(\int_{i\delta}^{i\Lambda}\frac{\dd{z}}{2\pi i}4\ev{\bar{c}(\bar{z})\partial c(-L_1)c(-L_1-L_2)}_{C_s}\ev{\partial X^0(z)\bar{\partial}X^0(\bar{z})e^{ik_1\cdot X}(z_1)e^{ik_2\cdot X}(z_2)}_{C_s}\right.\\
    &\qquad\qquad\qquad-\int_{-i\delta}^{-i\Lambda}\frac{\dd{\bar{z}}}{2\pi i}4\ev{c(z)\partial c(-L_1)c(-L_1-L_2)}_{C_s}\ev{\partial X^0(z)\bar{\partial}X^0(\bar{z})e^{ik_1\cdot X}(z_1)e^{ik_2\cdot X}(z_2)}_{C_s}\\
    &\qquad\qquad\qquad\left.+\frac{1}{2\pi\delta}\ev{c(0)\partial c(-L_1)c(-L_1-L_2)}_{C_s}\ev{e^{ik_1\cdot X}(z_1)e^{ik_2\cdot X}(z_2)}_{C_s}\right)\\
    &\qquad+i\sqrt{h}\lim_{(\Lambda,\delta)\to(\infty,0)}\left(\int_{i\delta}^{i\Lambda}\frac{\dd{z}}{2\pi i}4\ev{\bar{c}(\bar{z})c(-L_1)c(-L_1-L_2)}_{C_s}\ev{\partial X^0(z)\bar{\partial}X^0(\bar{z})\partial X^0(z_1)e^{ik_1\cdot X}(z_1)e^{ik_2\cdot X}(z_2)}_{C_s}\right.\\
    &\qquad\qquad\qquad-\int_{-i\delta}^{-i\Lambda}\frac{\dd{\bar{z}}}{2\pi i}4\ev{c(z)c(-L_1)c(-L_1-L_2)}_{C_s}\ev{\partial X^0(z)\bar{\partial}X^0(\bar{z})\partial X^0(z_1)e^{ik_1\cdot X}(z_1)e^{ik_2\cdot X}(z_2)}_{C_s}\\
    &\qquad\qquad\qquad\left.+\frac{1}{2\pi\delta}\ev{c(0)c(-L_1)c(-L_1-L_2)}_{C_s}\ev{\partial X^0(z_1)e^{ik_1\cdot X}(z_1)e^{ik_2\cdot X}(z_2)}_{C_s}\right),\label{piece2}
\end{align}
and
\begin{align}
    &\Tr(\chi e^{-L_1K}e^{ik_1\cdot X}e^{-L_2K}ce^{-L_3K}Qe^{ik_2\cdot X}e^{-L_4K})\\
    &=-h\Tr(\chi e^{-L_1K}e^{ik_1\cdot X}e^{-L_2K}ce^{-L_3K}\partial ce^{ik_2\cdot X}e^{-L_4K})-i\sqrt{h}\Tr(\chi e^{-L_1K}e^{ik_1\cdot X}e^{-L_2K}ce^{-L_3K}c\partial X^0e^{ik_2\cdot X}e^{-L_4K})\\
    &=-h\lim_{(\Lambda,\delta)\to(\infty,0)}\left(\int_{i\delta}^{i\Lambda}\frac{\dd{z}}{2\pi i}4\ev{\bar{c}(\bar{z})c(-L_1-L_2)\partial c(-L_1-L_2-L_3)}_{C_s}\ev{\partial X^0(z)\bar{\partial}X^0(\bar{z})e^{ik_1\cdot X}(z_1)e^{ik_2\cdot X}(z_2)}_{C_s}\right.\\
    &\qquad\qquad\qquad-\int_{-i\delta}^{-i\Lambda}\frac{\dd{\bar{z}}}{2\pi i}4\ev{\bar{c}(\bar{z})c(-L_1-L_2)\partial c(-L_1-L_2-L_3)}_{C_s}\ev{\partial X^0(z)\bar{\partial}X^0(\bar{z})e^{ik_1\cdot X}(z_1)e^{ik_2\cdot X}(z_2)}_{C_s}\\
    &\qquad\qquad\qquad\left.+\frac{1}{2\pi\delta}\ev{c(0)c(-L_1-L_2)\partial c(-L_1-L_2-L_3)}_{C_s}\ev{e^{ik_1\cdot X}(z_1)e^{ik_2\cdot X}(z_2)}_{C_s}\right)\\
    &-i\sqrt{h}\lim_{(\Lambda,\delta)\to(\infty,0)}\left(\int_{i\delta}^{i\Lambda}\frac{\dd{z}}{2\pi i}4\ev{\bar{c}(\bar{z})c(-L_1-L_2)c(-L_1-L_2-L_3)}_{C_s}\ev{\partial X^0(z)\bar{\partial}X^0(\bar{z})e^{ik_1\cdot X}(z_1)\partial X^0(z_2)e^{ik_2\cdot X}(z_2)}_{C_s}\right.\\
    &\qquad\qquad-\int_{-i\delta}^{-i\Lambda}\frac{\dd{\bar{z}}}{2\pi i}4\ev{\bar{c}(\bar{z})c(-L_1-L_2)c(-L_1-L_2-L_3)}_{C_s}\ev{\partial X^0(z)\bar{\partial}X^0(\bar{z})e^{ik_1\cdot X}(z_1)\partial X^0(z_2)e^{ik_2\cdot X}(z_2)}_{C_s}\\
    &\qquad\qquad\left.+\frac{1}{2\pi\delta}\ev{c(0)c(-L_1-L_2)c(-L_1-L_2-L_3)}_{C_s}\ev{e^{ik_1\cdot X}(z_1)\partial X^0(z_2)e^{ik_2\cdot X}(z_2)}_{C_s}\right).\label{piece3}
\end{align}
To satisfy \eqref{cond:3}, the sum of \eqref{piece1}, \eqref{piece2} and \eqref{piece3} has to coincide with \eqref{EllwoodX0} and we examine it order by order in $h$. Because of
\begin{align}
    \ev{\partial X^0(z)\bar{\partial}X^0(\bar{z})}_{C_s}&\sim\order{h^0},\\
    \ev{\partial X^0(z)e^{ik_m\cdot X(z_m)}}_{C_s}&\sim\order{h^{1/2}},
\end{align}
we obtain
\begin{align}
    \Tr'(\chi Q\Psi)=\Tr'(\chi Q\Psi_{L_1+L_2,L_3+L_4})+\order{h}.
\end{align}
Because the first term can be expressed
\begin{align}
    \Tr'\qty(\chi Q\Psi_{L_1+L_2,L_3+L_4})=\Tr'\qty(\frac{2}{\pi i}c\bar{c}\partial X^0\bar{\partial}X^0\Psi_{L_1+L_2,L_3+L_4}),
\end{align}
the higher-order terms in $h$ has to vanish. We focus on the term at $\order{h^2}$. It is given by
\begin{align}
    &\lim_{(\Lambda,\delta)\to(\infty,0)}\sum_{\substack{1\le o\le2\\1\le m\le2}}\left[-4h\int_{i\delta}^{i\Lambda}\frac{\dd{z}}{2\pi i}\left(\qty(\ev{\bar{c}(\bar{z})\partial c(-L_1)c(-L_1-L_2)}_{C_s}-\ev{\bar{c}(\bar{z})c(-L_1-L_2)\partial c(-L_1-L_2-L_3)}_{C_s})\right.\right.\\
    &\qquad\left.\times\ev{\partial X^0(z)e^{ik_m\cdot X(z_m)}}_{C_s}\ev{\bar{\partial}X^0(\bar{z})e^{ik_o\cdot X(z_o)}}_{C_s}\right)\\
    &\quad+4h\int_{-i\delta}^{i\Lambda}\frac{\dd{\bar{z}}}{2\pi i}\left(\qty(\ev{c(z)\partial c(-L_1)c(-L_1-L_2)}_{C_s}-\ev{c(z)c(-L_1-L_2)\partial c(-L_1-L_2-L_3)}_{C_s})\right.\\
    &\qquad\left.\left.\times\ev{\partial X^0(z)e^{ik_m\cdot X(z_m)}}_{C_s}\ev{\bar{\partial}X^0(\bar{z})e^{ik_o\cdot X(z_o)}}_{C_s}\right)\right]\\
    &=-4h^2\lim_{(\Lambda,\delta)\to(\infty,0)}\sum_{\substack{1\le o\le2\\1\le m\le2}}\operatorname{sgn}(k_o^0k_m^0)\cos\frac{\pi(L_2+L_3)}{s}\cos\frac{\pi}{s}z_m\cos\frac{\pi}{s}z_o\\
    &\quad\left[\int_{i\delta}^{i\Lambda}\frac{\dd{z}}{2\pi i}\frac{\cos\frac{2\pi(\bar{z}+L_1)+\pi(L_2+L_3)}{s}-\cos\frac{\pi(L_2-L_3)}{s}}{\sin\frac{\pi}{s}(z_o-\bar{z})\sin\frac{\pi}{s}(z_m-z)\cos\frac{\pi}{s}z\cos\frac{\pi}{s}\bar{z}}-\int_{-i\delta}^{-i\Lambda}\frac{\dd{\bar{z}}}{2\pi i}\frac{\cos\frac{2\pi(z+L_1)+\pi(L_2+L_3)}{s}-\cos\frac{\pi(L_2-L_3)}{s}}{\sin\frac{\pi}{s}(z_o-\bar{z})\sin\frac{\pi}{s}(z_m-z)\cos\frac{\pi}{s}z\cos\frac{\pi}{s}\bar{z}}\right].\label{result:chi}
\end{align}
In Figure \ref{fig:chi}, we present numerical result.
It is clear that \eqref{result:chi} is nonzero. Therefore the equation \eqref{cond:3} does not hold for the string field involving $X^0$ variables.

\section{Examination of condition (\ref{cond:1a}) and (\ref{cond:1b})\label{cond:1}}
We would like to examine the conditions \eqref{cond:1a} and \eqref{cond:1b}. In this appendix, we consider more general condition
\begin{align}
    \Tr(\mathcal{G}\Psi_1\Psi_2\dots\Psi_N)+\Tr(\Psi_1\mathcal{G}\Psi_2\dots\Psi_N)+\dots+\Tr(\Psi_1\Psi_2\dots\mathcal{G}\Psi_N)=\Tr(\Psi_1\Psi_2\dots\Psi_N)
\end{align}
than \eqref{cond:1a} and \eqref{cond:1b}. We show that the above equation holds for arbitrary string fields $\Psi_i$ which are constructed only by $K,B,c$ and matter operators.

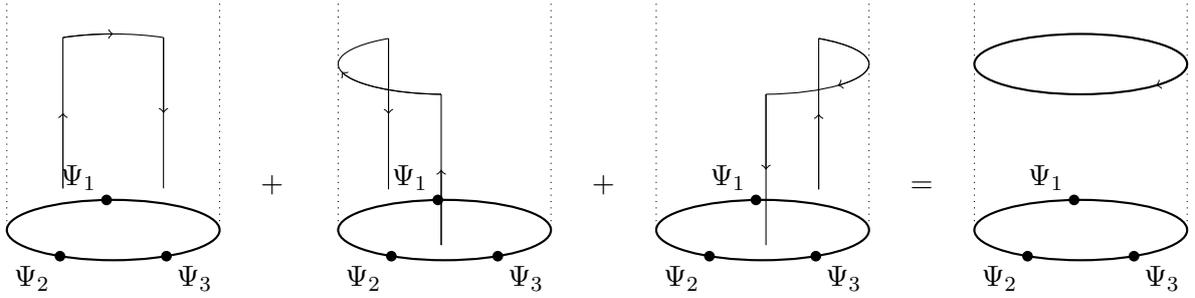
\begin{figure}
    \centering
    \begin{tikzpicture}[baseline=0.5cm]
        \draw[thick] (0,0) ellipse (1.4 and 0.4);
        \fill ({1.4*cos(pi*(1/2+0.02) r)},{0.4*sin(pi*(1/2+0.02) r)}) circle (2pt) node[above left] {$\Psi_1$};
        \fill ({1.4*cos(pi*4/3 r)},{0.4*sin(pi*4/3 r)}) circle (2pt) node[below left] {$\Psi_2$};
        \fill ({1.4*cos(-pi/3 r)},{0.4*sin(-pi/3 r)}) circle (2pt) node[below right] {$\Psi_3$};
        %
        \draw ({1.4*cos(pi*(1/3+0.01) r)},{0.4*sin(pi*(1/3+0.01) r)+0.2})--++(0,2);
        \draw ({1.4*cos(pi*(1/3+0.01) r)},{0.4*sin(pi*(1/3+0.01) r)+2.2}) arc (pi*(1/3+0.01) r:pi*(2/3-0.01) r:1.4 and 0.4);
        \draw ({1.4*cos(pi*(2/3-0.01) r)},{0.4*sin(pi*(2/3-0.01) r)+0.2})--++(0,2);
        \draw[->] ({1.4*cos(pi*(1/3+0.01) r)},{0.4*sin(pi*(1/3+0.01) r)+2.2})--++(0,-1);
        \draw[->] ({1.4*cos(pi*(2/3-0.01) r)},{0.4*sin(pi*(2/3-0.01) r)+2.2}) arc (pi*(2/3-0.01) r:pi/2 r:1.4 and 0.4);
        \draw[->] ({1.4*cos(pi*(2/3-0.01) r)},{0.4*sin(pi*(2/3-0.01) r)+0.2})--++(0,1);
        \draw[dotted] (1.4,3)--(1.4,0);
        \draw[dotted] (-1.4,3)--(-1.4,0);
    \end{tikzpicture}
    \hspace{1em}+\hspace{1em}
    \begin{tikzpicture}[baseline=0.5cm]
        \draw[thick] (0,0) ellipse (1.4 and 0.4);
        \fill ({1.4*cos(pi*(1/2+0.02) r)},{0.4*sin(pi*(1/2+0.02) r)}) circle (2pt) node[above left] {$\Psi_1$};
        \fill ({1.4*cos(pi*4/3 r)},{0.4*sin(pi*4/3 r)}) circle (2pt) node[below left] {$\Psi_2$};
        \fill ({1.4*cos(-pi/3 r)},{0.4*sin(-pi/3 r)}) circle (2pt) node[below right] {$\Psi_3$};
        %
        \draw ({1.4*cos(pi*(2/3+0.01) r)},{0.4*sin(pi*(2/3+0.01) r)+0.2})--++(0,2);
        \draw ({1.4*cos(pi*(2/3+0.01) r)},{0.4*sin(pi*(2/3+0.01) r)+2.2}) arc (pi*(2/3+0.01) r:pi*(3/2-0.01) r:1.4 and 0.4);
        \draw ({1.4*cos(pi*(3/2-0.01) r)},{0.4*sin(pi*(3/2-0.01) r)+0.2})--++(0,2);
        \draw[->] ({1.4*cos(pi*(2/3+0.01) r)},{0.4*sin(pi*(2/3+0.01) r)+2.2})--++(0,-1);
        \draw[->] ({1.4*cos(pi*(3/2-0.01) r)},{0.4*sin(pi*(3/2-0.01) r)+2.2}) arc (pi*(3/2-0.01) r:pi*13/12 r:1.4 and 0.4);
        \draw[->] ({1.4*cos(pi*(3/2-0.01) r)},{0.4*sin(pi*(3/2-0.01) r)+0.2})--++(0,1);
        \draw[dotted] (1.4,3)--(1.4,0);
        \draw[dotted] (-1.4,3)--(-1.4,0);
    \end{tikzpicture}
    \hspace{1em}+\hspace{1em}
    \begin{tikzpicture}[baseline=0.5cm]
        \draw[thick] (0,0) ellipse (1.4 and 0.4);
        \fill ({1.4*cos(pi*(1/2+0.02) r)},{0.4*sin(pi*(1/2+0.02) r)}) circle (2pt) node[above left] {$\Psi_1$};
        \fill ({1.4*cos(pi*4/3 r)},{0.4*sin(pi*4/3 r)}) circle (2pt) node[below left] {$\Psi_2$};
        \fill ({1.4*cos(-pi/3 r)},{0.4*sin(-pi/3 r)}) circle (2pt) node[below right] {$\Psi_3$};
        %
        \draw ({1.4*cos(pi*(3/2+0.01) r)},{0.4*sin(pi*(3/2+0.01) r)+0.2})--++(0,2);
        \draw ({1.4*cos(pi*(3/2+0.01) r)},{0.4*sin(pi*(3/2+0.01) r)+2.2}) arc (pi*(3/2+0.01) r:pi*(7/3-0.01) r:1.4 and 0.4);
        \draw ({1.4*cos(pi*(7/3-0.01) r)},{0.4*sin(pi*(7/3-0.01) r)+0.2})--++(0,2);
        \draw[->] ({1.4*cos(pi*(3/2+0.01) r)},{0.4*sin(pi*(3/2+0.01) r)+2.2})--++(0,-1);
        \draw[->] ({1.4*cos(pi*(7/3-0.01) r)},{0.4*sin(pi*(7/3-0.01) r)+2.2}) arc (pi*(7/3-0.01) r:pi*7/4 r:1.4 and 0.4);
        \draw[->] ({1.4*cos(pi*(7/3-0.01) r)},{0.4*sin(pi*(7/3-0.01) r)+0.2})--++(0,1);
        \draw[dotted] (1.4,3)--(1.4,0);
        \draw[dotted] (-1.4,3)--(-1.4,0);
    \end{tikzpicture}
    \hspace{1em}=\hspace{1em}
    \begin{tikzpicture}[baseline=0.5cm]
        \draw[thick] (0,0) ellipse (1.4 and 0.4);
        \fill ({1.4*cos(pi*(1/2+0.02) r)},{0.4*sin(pi*(1/2+0.02) r)}) circle (2pt) node[above left] {$\Psi_1$};
        \fill ({1.4*cos(pi*4/3 r)},{0.4*sin(pi*4/3 r)}) circle (2pt) node[below left] {$\Psi_2$};
        \fill ({1.4*cos(-pi/3 r)},{0.4*sin(-pi/3 r)}) circle (2pt) node[below right] {$\Psi_3$};
        %
        \draw[thick] (0,2.2) ellipse (1.4 and 0.4);
        \draw[->] ({1.4*cos(pi*(7/3-0.01) r)},{0.4*sin(pi*(7/3-0.01) r)+2.2}) arc (pi*(7/3-0.01) r:pi*7/4 r:1.4 and 0.4);
        \draw[dotted] (1.4,3)--(1.4,0);
        \draw[dotted] (-1.4,3)--(-1.4,0);
    \end{tikzpicture}
    \caption{The contour of $\displaystyle\sum_{j=1}^N\mathcal{G}(L_j,\Lambda,\delta)$ at the case of $N=3$}\label{figure:contourcancel}
\end{figure}

If we represent $\Psi_i$ by a Laplace transform, the left-hand side can be written by
\begin{align}
    &\Tr(\mathcal{G}\Psi_1\Psi_2\dots\Psi_N)+\Tr(\Psi_1\mathcal{G}\Psi_2\dots\Psi_N)+\dots+\Tr(\Psi_1\Psi_2\dots\mathcal{G}\Psi_N)\\
    &=\prod_{i=1}^N\qty(\int_0^\infty\dd{L_i})\lim_{(\Lambda,\delta)\to(\infty,0)}\ev{e^{sK}\mathcal{G}(L_1,\Lambda,\delta)e^{-L_1K}\psi(L_1)e^{-L_2K}\psi(L_2)\dots e^{-L_NK}\psi(L_N)}_{C_s}\\
    &\qquad+\prod_{i=1}^N\qty(\int_0^\infty\dd{L_i})\lim_{(\Lambda,\delta)\to(\infty,0)}\ev{e^{sK}e^{-L_1K}\psi(L_1)\mathcal{G}(L_2,\Lambda,\delta)e^{-L_2K}\psi(L_2)\dots e^{-L_NK}\psi(L_N)}_{C_s}\\
    &\qquad\vdots\\
    &\qquad+\prod_{i=1}^N\qty(\int_0^\infty\dd{L_i})\lim_{(\Lambda,\delta)\to(\infty,0)}\ev{e^{sK}e^{-L_1K}\psi(L_1)e^{-L_2K}\psi(L_2)\dots\mathcal{G}(L_N,\Lambda,\delta)e^{-L_NK}\psi(L_N)}_{C_s},
\end{align}
where $s$ is given by
\begin{align}
    s=\sum_{i=1}^NL_i.
\end{align}
Using the cyclicity of the cylinder, the vertical part of the contours cancel each other (Figure \ref{figure:contourcancel}).
Additionally, using \eqref{formula:deltafactorize}, we obtain
\begin{align}
    &\Tr(\mathcal{G}\Psi_1\Psi_2\dots\Psi_N)+\Tr(\Psi_1\mathcal{G}\Psi_2\dots\Psi_N)+\dots+\Tr(\Psi_1\Psi_2\dots\mathcal{G}\Psi_N)\\
    &=\prod_{i=1}^N\qty(\int_0^\infty\dd{L_i})\lim_{(\Lambda,\delta)\to(\infty,0)}\ev{\mathcal{G}(s,\Lambda,\delta)}_{C_s}\ev{e^{sK}e^{-L_1K}\psi(L_1)e^{-L_2K}\psi(L_2)\dots e^{-L_NK}\psi(L_N)}_{C_s}\\
    &=\Tr(\Psi_1\Psi_2\dots\Psi_N).
\end{align}
This is what we wanted to show. Therefore it is clear that \eqref{cond:1a} and \eqref{cond:1b} hold for the solutions which are constructed only by $K,B,c$ and matter operators.
\printbibliography
\end{document}